\RequirePackage{fix-cm}
\documentclass[smallextended]{svjour3}       
\smartqed  
\usepackage{graphicx}
\usepackage{siunitx} 
\usepackage{graphicx} 
\usepackage{enumerate}
\usepackage{subfigure}
\usepackage{cite}
\usepackage{dsfont}
\usepackage{amsfonts}
\usepackage[utf8]{inputenc}
\usepackage{algorithm}
\usepackage[noend]{algpseudocode}
\usepackage{times} 
\usepackage{amsmath} 
\usepackage{epsfig,wrapfig} 
\usepackage{blindtext}

\newtheorem{Claim}{Claim}

\newtheorem{Remark}{Remark}

\newtheorem{assumption}{Assumption}

\usepackage{color}
 \journalname{ }

\begin{document}

\title{From Instantaneous Schedulability to Worst Case Schedulability: A Significant Moment Approach}


\author{Ningshi Yao         \and
        Fumin Zhang 
}


\institute{Ningshi Yao \at
              Department of Electrical and Computer Engineering, George Mason University, Fairfax, VA 22033\\
              \email{nyao4@gmu.edu}           
           \and
           Fumin Zhang \at
              School of Electrical and Computer Engineering, Georgia Institute of Techonology, Atlanta, GA 30308\\
              \email{fumin@gatech.edu}           
}

\date{Received: date / Accepted: date}

\maketitle

\begin{abstract}
The method of significant moment analysis has been employed to derive instantaneous schedulability tests for real-time systems. However, the instantaneous schedulability can only be checked within a finite time window. On the other hand, worst-case schedulability  guarantees schedulability of systems for infinite time. This paper derives the classical worst-case schedulability conditions for preemptive periodic systems starting from instantaneous schedulability, hence unifying the two notions of schedulability. The results provide a rigorous justification on the critical time instants being the worst case for scheduling of preemptive periodic systems. The paper also show that the critical time instant is not the only worst case moments.

\keywords{worst-case schedulability \and instantaneous schedulability \and critical instant analysis}
\end{abstract}

\section{Introduction}
\label{intro}

Research on schedulability was initiated by the work \cite{liu1973scheduling}. In this foundational work, Liu and Layland considered a real-time computing system with the following assumptions: (1) all tasks are periodic, preemptive, and synchronized; (2) all tasks have their relative deadlines equal to their periods. They introduced an idea of {\it critical instant analysis} to study the worst case when all scheduling tasks are requested at the same time. At the critical time instant, a task set will endure the longest response time. Hence, a task set will be schedulable if they are schedulable at the critical time instant. Consider a set of $N$ tasks where $U_i$ represents the processor utilization for  the individual task $i$. Then   under the rate-monotonic scheduling (or RMS) algorithm, a sufficient condition for schedulability is that  the total processor utilization satisfies $\sum_{i=1}^{N}U_i \!\leq\! N(2^{1/N}\!-\!1)$. Meanwhile, under the earliest deadline first (or EDF) algorithm, the sufficient condition for schedulability is that the total processor utilization satisfies $\sum_{i=1}^{N}U_i \!\leq\! 1$. 

By extending the critical instant analysis,  extensive research has been conducted to improve the schedulability tests in Liu and Layland's work. This result was generalized to all periodic tasks with arbitrary release times in \cite{jeffay1989analysis}. The sufficient and necessary schedulability tests for task sets with arbitrary deadlines were developed in \cite{lehoczky1990fixed,tindell1994extendible}. Bini et al. \cite{bini2003rate} derived a hyperbolic bound for tasks scheduled under RMS. The hyperbolic bound is less pessimistic than Liu and Layland's utilization bound. Lehoczky et al. \cite{lehoczky1989rate} showed that the average processor utilization for a large set of randomly chosen tasks schedulable under RMS, is approximately $88\%$. Abdelzaher et al \cite{abdelzaher2004utilization} relaxed the periodic restriction on tasks and derived an utilization bound for nonperiodic tasks. In work \cite{jeffay1991non}, the authors presented a necessary and sufficient schedulabilty condition for non-preemptive systems in discrete time under non-preemptive EDF scheduling. These tests were
extended by George et al. in 1996, to the general case of
sporadic task sets with arbitrary deadlines \cite{george:inria-00073732}. In this paper, George et al. also derived the sufficient and necessary condition for fixed priority non-preemptive scheduling of arbitrary deadline
task sets. In \cite{marouf2012schedulability}, the schedulability analysis for a combination of non-preemptive periodic tasks and preemptive sporadic tasks under fixed priority scheduling was presented. 
Indeed, the critical instant analysis and the associated worst case schedulability tests have a proven track record.

 The critical instants are just a special case of a broader class of time instants that carry interesting information about the scheduled behaviors of a task set. A class of such time instants are named as {\it significant moments} in  \cite{shi2013predicting}.   
 Using the method of {\it significant moment analysis},
 a timing model was derived to capture the dynamic behaviors of a set of tasks under real-time scheduling with time dependent  processor time-occupancy and dynamic deadlines. This approach turns out to be useful for several applications. In work \cite{shi2015model}, a timing model for non-preemptive periodic tasks was developed and applied to the control area network (CAN). More advanced timing models for both preemptive and non-preemptive tasks are derived and applied to energy management in micro-grids \cite{SYZ16} and traffic scheduling at an intersection \cite{yao2018resolving, yao2019contention, yao2021contention}. Timing models have also been derived for discrete-time systems arising in the allocation of human attention for multiple robots \cite{wang2015dynamic, yao2020optimal}. Moreover, the timing model enables an event-triggered model predictive control approach that are able to solve the control and scheduling co-design problem that has been a lasting theme of research in real-time systems \cite{yao2020contention, yao2020optimal, yao2020intersection}.
 
 The significant moment analysis, and the resulting timing models for task sets with different properties, have been leveraged for schedulability analysis. Due to the dynamic nature of the timing models, the notion of {\it instantaneous schedulability} was introduced in \cite{shi2015model}. The condition for instantaneous schedulability  was proved to be both sufficient and necessary. Hence, by checking the instantaneous schedulability of all significant moments within a finite time interval, the scheduability of a task sets can be guaranteed {\it within that time interval}. This finite time interval schedulability guarantee is acceptable for many applications, however, it cannot guarantee schedulability for an infinite time like the worst case schedulability do. Hence, there is a theoretical gap between the instantaneous shedulability and the worst case schedulability that needs to be fulfilled. 

In this paper, we show that the two notions of schedulability can indeed be unified. We revisit Liu and Layland's canonical task sets and utilize the significant moment analysis to derive the same set of worst case schedulability conditions under RMS. The timing models for the task sets are derived and all significant moments are characterized. We will show that the critical instant analyzed in \cite{liu1973scheduling} is not the only worst case scenario. In other words, there exist other significant moments that are equally ``bad".  This new discovery has not been seen in the literature. Then, we show that the instantaneous schedulability conditions checked in all worst cases lead to the sufficient condition of Liu and Layland. These new insights may help researchers to derive worst case schedulability and instantaneous schedulability conditions for more challenging task sets, such as a non-preemptive task set. 



This paper is organized as follows. In Section \ref{SecSMA}, we recap the analytical timing model and instantaneous schedulability. In Section \ref{Sec_Timing_Model:InfiniteTiming Schedulability}, we present the method to derive infinite-time schedulability for preemptive and periodic tasks. Section \ref{Conclusion} presents the conclusion.

\section{Significant Moment Analysis and Timing Model}
\label{SecSMA}
 In this section, we present the characteristics of real-time scheduling tasks and briefly recap the analytical timing models and instantaneous schedulability test condition developed in \cite{shi2015model}.

\label{Formulation}
\subsection{Task Characteristics}
\begin{figure}[t]
\centering
\includegraphics[width=4.6in]{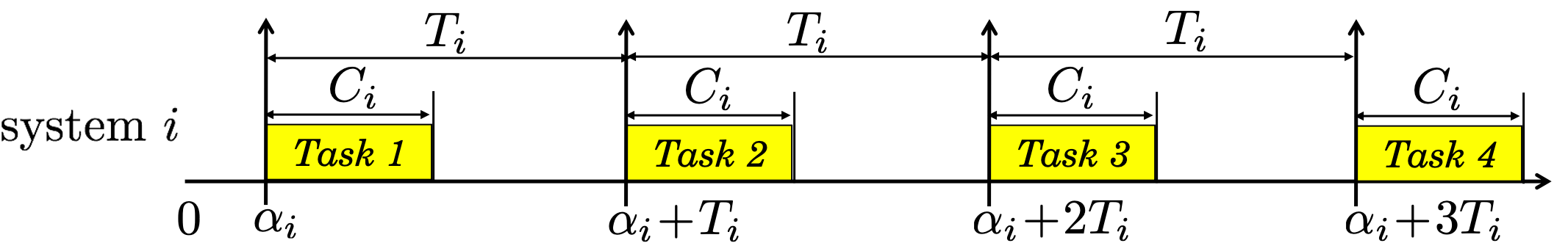}
\caption{An example of tasks from system $i$ when no contention occurs. The yellow rectangles represents the amount of time system $i$ needs to occupy the share resource.}
\label{Timing_Characteristics}
\end{figure}

In real-time scheduling, each system has a sequence of tasks which require to utilize a share resource. The time instant when system $i$ makes the first request to use the shared resource is denoted by $\alpha_i$. The following assumptions are made about the real-time systems.
\begin{assumption}\label{assumption_1}
The requests to use the shared resource for all tasks of system $i$ are periodic, with a constant interval $T_i$ between requests.
\end{assumption}
\begin{assumption}\label{assumption_2}
The amount of time that tasks from system $i$ require to occupy the resource is also a constant, denoted by $C_i$. The constant satisfies $C_i<T_i$ for all $i$.
\end{assumption}
\begin{assumption}\label{assumption_3}
For systems $i$ and $j$ where $i\!<\!j$ , we have $T_i<T_j$.
\end{assumption}
The timing characteristics of tasks from system $i$ are shown in Figure \ref{Timing_Characteristics}. The request starting time of the $k$th task is $\alpha_i\!+\!(k\!-\!1)T_i$. If there are no contentions among real-time systems, then the completion time of the $k$th task is $\alpha_i\!+\!(k\!-\!1)T_i\!+\!C_i$. If multiple systems request to use the shared resource at the same time, a contention occurs and priorities are needed to resolve contentions. Priorities can be determined by different scheduling methods and the completion times of the tasks of lower prioritized systems are delayed by the higher prioritized systems.

\subsection{Timing States}\label{TimingModelSection}
To describe the timing behavior at any time $t$, the timing state variable are defined $Z(t)\!=\!(D(t),R(t),O(t))$ as follows.

The deadline variable is $D(t)\!=\!(d_1(t),...,d_i(t),...,d_N(t))$, where $d_i(t)$ denotes how long after time $t$ the next task of system $i$ will be generated.
The remaining time variable is $R(t)\!=\!(r_1(t),...,r_i(t),...,r_N(t))$, where $r_i(t)$ is the remaining time after time $t$ that is required to complete the most recently generated task of system $i$. And the dynamic response time variable is $O(t)\!=\!(o_1(t),...,o_i(t),...,o_N(t))$, where $o_i(t)$ denotes the length of time from the most recent request from system $i$ to the minimum of (a) the time when the most recent request from system $i$ is completed and (b) the current time $t$. For example, if a task of system $i$ starts the request to use the shared resource at time $t$, then we have $d_i(t)\!=\!T_i$ and $r_i(t)\!=\!C_i$. 

A priority assignment is a tuple $\mathbf {P}(t)\!=\!(p_1(t),...,p_i(t),...,p_N(t))\!\in\! \mathcal{P}(\!\{1,...,N\}\!)$, where $p_i(t)$ is the priority assigned to system $i$ at time $t$ and such that for each $i$ and $j$ in $\{1,...,N\}$, we have $p_i(t)\!<\!p_j(t)$ if and only if system $i$ is assigned higher priority than system $j$ at time $t$. For each $t\!\in\! [t_0,t_f]$, the value of $p_i(t)$ is a positive integer in $\{1,\ldots,N\}$, such that $p_i(t)\!\neq\! p_j(t)$ if $i\!\neq\! j$. When a contention occurs, only the system with the smallest $p_i$ will be granted access. The completion times of lower prioritized tasks are delayed by the higher prioritized systems. The time delay at $t$ can be computed using the timing state as $\max(0,o_i(t)\!-\!C_i)$. 
In a real-time system, it is required that each task must be completed before its deadline, in order for the system to be schedulable. In this paper, the deadline for the $k$th task is defined to be the time instant when the next task of system $i$ is generated, i.e., $\alpha_i\!+\!kT_i$. 
\begin{Remark}
The {\it critical time instant} in  \cite{liu1973scheduling} is defined as the time instants when all systems start the request to use the shared resource at the same time. Using the definition of timing states, a {\it critical time instant} is a time $t$ when $d_i(t)\!=\!T_i$ and $r_i(t)\!=\!C_i$ are satisfied for all $i$.
\end{Remark}

\subsection{Timing Model}\label{TimingModelSection}
The timing model, which is the evolution rule for $Z(t)$ within a time interval $[t_0,t_f]$ given the timing characteristics $(\alpha_i,C_i,T_i)_{i=1,...,N}$ and priority assignment $\mathbf{P}(t)$, consists of a set of analytical algebraic and differential equations that can account for time-varying priorities and interruption of access to the resource by higher priority tasks. The detailed equations can be found in \cite{shi2013predicting, yao2020contention}. One benefit of this timing model is a necessary and sufficient condition for the schedulability of all tasks in a finite time interval.
\begin{definition}
Tasks from system $i$ are instantaneously schedulable during time interval $[t_1, t_2]$ if $r_i(t)\!\le\! d_i(t)$ for any $t\!\in\![t_1, t_2]$.
\end{definition}
If tasks from system $i$ are instantaneously schedulable for all time $t\!\in\![t_1, t_2]$, then all the remaining time variables are less than or equal to the deadline variables, meaning all the deadlines are met, then tasks from system $i$ are schedulable.

It is also discovered that even though the real-time systems evolve continuously in time, there are certain moments that are more significant than others because the timing states change discontinuously at these moments due to whether access to the shared resource is granted or not. The first significant moment $t_0$ is defined to be starting time $0$ and $d_i(t_0)\!=\!\alpha_i$. The significant moments for periodic and preemptive systems can be computed
\begin{align}\label{equation:preemptive}
t_{w+1}\!-\!t_{w}\!=\!\min \left \{d_1(t_w),...,d_N(t_w), t_2\!-\!t_{w}\right \}
\end{align}
where $t_{w}$ denotes the significant moment and $w=0,1,2,...$.
The values of timing states may have jumps at significant moments and evolve continuously between two successive significant moments. Here, we emphasize two properties, which will be used in later sections. Within two successive significant moments, i.e., $t\!\in\!(t_w, t_{w+1})$, the changing rate of deadline variable is $\dot{d}_i(t)\!=\!-\!1$. Therefore, the equation for $d_i(t)$ is $d_i(t)=d_i(t_w)-(t\!-\!t_w)$. For the remaining time variable, $\dot{r}_i(t)\!=\!-\!1$ if system $i$ is occupying the shared resource and $\dot{r}_i(t)\!=\!0$ if system $i$ is interrupted by a higher prioritized system at $t$. Then the equation for the remaining time can be derived as $r_i(t)\!=\!{\rm max}\! \left\{\!0, r_i(t_w)\!-\!{\rm max}\! \left\{0, t\!-\!t_w\!-\! \! \! \!  \! {\sum\limits_{j\in \mathrm{HP}(i)}\!\!\! r_j(t_w)}\!\right\}\!\right\}$ where ${\rm HP}(i)$ denotes a set of indices of systems which have a higher priority than system $i$.

Based on the significant moments, the instantaneous schedulability only needs to be checked for finitely many times.
\begin{corollary}\label{ins_schedulability}
Tasks are instantaneously schedulable at any time $t\!\in\![t_1, t_2]$ if $r_i(t_w^-)\!\le\! d_i(t_w^-)$ is satisfied at any significant moment $t_w\!\in\![t_1, t_2]$, where $t_w^-$ denotes the limit from the left.
\end{corollary}
This corollary implies that if we want to check whether a system is schedulable or not, all we need to do is to check whether it is instantaneously schedulable right before the significant moments. The detailed proof is included in \cite{shi2015model}. The instantaneous schedulability condition is sufficient and necessary and only requires to check for finite many times within a finite time interval. 

The difficult part occurs when the time interval $[t_1, t_2]$ is infinitely long, which makes the condition in \textbf{Corollary \ref{ins_schedulability}} incomputable because we cannot check this condition for infinitely many times. The schedulability for infinite-time horizon is addressed by the worst-case schedulability in real-time scheduling. We will derive the worst-case schedulability conditions using the instantaneous schedulability conditions for preemptive and periodic tasks in the next section.

\section{Worst-case Schedulability}\label{Sec_Timing_Model:InfiniteTiming Schedulability}
In this section, we reproduce the worst case schedulability conditions as in \cite{liu1973scheduling} using  significant moment analysis. We investigate a task set from $N$ periodic and preemptive systems. First, the special case of $N\!=\!2$ is considered, then the results are generalized to any value of $N$ as a positive integer.  

\begin{figure}[t]
\centering
\includegraphics[width=3.8in]{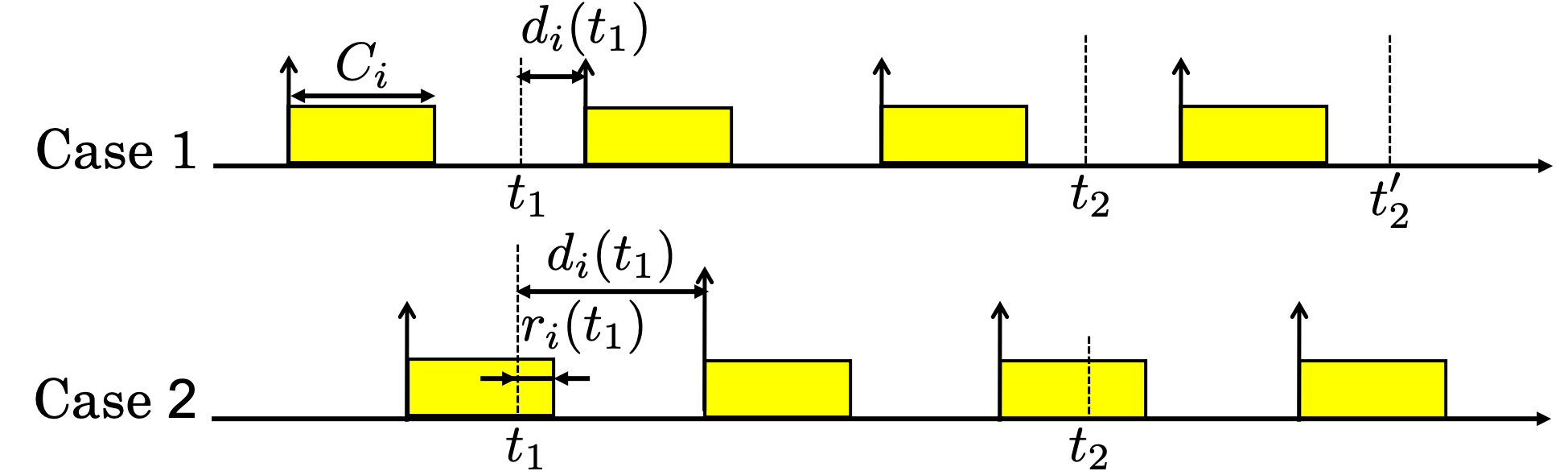}
\caption{Illustration of time-occupancy. The yellow rectangles represent the time occupied by system $i$.}
\label{Occupation_Time}
\end{figure}

We first introduce a definition of time-occupancy for tasks of system $i$ within an arbitrary time interval $[t_1,t_2]$.
\begin{definition}\label{Def_occupancy}
The {\it time-occupancy}, denoted as ${\rm OP}_i(t_1,t_2,d_i(t_1), r_i(t_1))$, is the total time occupied by system $i$ during a time interval $[t_1,t_2]$. The value of ${\rm OP}_i$ depends on the starting time of the time interval $t_1$, the ending time $t_2$, the deadline variable $d_i$ and the remaining time variable $r_i$ of system $i$ at time $t_1$. 
\end{definition}
By definition, the remaining time $r_i(t_1)$ represents the amount of time occupied by system $i$ within the incomplete period at the beginning of $[t_1,t_2]$. Therefore, ${\rm OP}_i$ depends on $r_i(t_1)$. It is also trivial why ${\rm OP}_i$ is a function of $t_1$ and $t_2$. As illustrated by Case $1$ in Figure \ref{Occupation_Time}, ${\rm OP}_i(t_1,t_2,d_i(t_1),r_i(t_1))\!=\!2C_i$. However, if the time interval is larger when $t_2$ changes to be $t_2'$, the time-occupancy ${\rm OP}_i(t_1,t_2',d_i(t_1),r_i(t_1))\!=\!3C_i$. The reason why the time-occupancy ${\rm OP}_i(t_1,t_2,d_i(t_1),r_i(t_1))$ is also a function of deadline variable is illustrate by Case $2$ in Figure \ref{Occupation_Time}. For the same time interval $[t_1,t_2]$, if the value of deadline variable $d_i(t_1)$ is increased, the time-occupancy ${\rm OP}_i(t_1,t_2,d_i(t_1),r_i(t_1))$ in this case is less than $2C_i$.

For periodic systems, we can provide a general formula for calculating ${\rm OP}_i$ as
\begin{align}\label{formula_general}
{\rm OP}_i(t_1,t_2,d_i(t_1), r_i(t_1))=r_i(t_1)+ N_c\,C_i +\tau_{{\rm e},i}    
\end{align}
where $N_c$ is the number of complete periods of system $i$ within $[t_1\!+\!d_i(t_1),t_2]$ and $\tau_{{\rm e},i}\!\ge\! 0$ is the time occupied by system $i$ within unfinished time interval at the end of $[t_1,t_2]$. This formula can be evaluated by knowing the timing states. Then for each case in the following arguments, all we need is to calculate the three terms in the formula. Using time-occupancy, schedulability can be characterized as
\begin{Claim}
System $i$ is schedulable within $[t_1, t_2]$ if $$\sum_{j\in {\rm HP(i)}}{\rm OP}_j(t_1, t_2, d_j(t_1), r_j(t_1))+{\rm OP}_i(t_1, t_2, d_i(t_1), r_i(t_1))\le t_2\!-\!t_1.$$       
\end{Claim}
\begin{Claim}
All systems are schedulable within $[t_1, t_2]$ if $$\sum_{i=1}^N{\rm OP}_i(t_1, t_2, d_i(t_1), r_i(t_1))\le t_2\!-\!t_1.$$
\end{Claim}
\begin{definition}
Given a specific priority assignment, the time $t_1$ when $d^*_i(t_1)$ is obtained for all $i\!\le\!N$ that maximizes $\sum_{i=1}^N{\rm OP}_i(t_1, t_2, d_i(t_1), r_i(t_1))$ is called a {\it worst-case moment} within the time interval $[t_1,t_2]$. The values $d^*_i(t_1)$ are called the \textit{worst-case deadline} for schedulability within interval $[t_1, t_2]$.
\end{definition}

If schedulability is maintained at these worst-case moments, then schedulability is maintained for all cases within the finite time interval. 

Given the \textbf{Assumption \ref{assumption_1}} and the timing model, if system $i$ is assigned with the highest priority, then $r_i(t_1)\!=\!{\rm max}\! \left\{0, r_i(t_w)\!-\!(t_1\!-\!t_w)\right\}$ where $t_w$ is the largest significant moment satisfying $d_i(t_w)\!=\!T_i$ and $t_w\!\le\! t_1$. Because $r_i(t_w)\!=\!C_i$ at time $t$ and $t_1\!-\!t_w\!=\!d_i(t_w)\!-\!d_i(t_1)$, we can rewrite the equation to compute the remaining time as 
\begin{align}\label{Remaining_Time_1}
r_i(t_w)\!=\!{\rm max}\! \left\{0,C_i\!-\!(d_i(t_w)\!-\!d_i(t_1))\right\}=\max\left\{0,d_i(t_1)\!-\!(T_i\!-\!C_i)\right\}.
\end{align} 
Therefore, if system $i$ has the highest priority, then its time-occupancy ${\rm OP}_i$ only depends on the deadline variable $d_i$. The result in (\ref{Remaining_Time_1}) will be used in the following sections.

\subsection{Schedulability for Two Systems}
It is well known that the critical instant is one of the worst-case moments. However, there has not been a systematic effort in the literature to define other possible worst-case moments.  Our first goal is to find out all possible worst-case moments.

\subsubsection{System $1$ has higher priority}\label{Section_two_systems_1}
For $N\!=\!2$, there are only two possible static priority assignments. The first priority assignment is that system $1$ always has higher priority than system $2$, i.e., $p_1(t)\!<\!p_2(t)$ for all $t$. Under this priority assignment, it is trivial that if $C_1\!<\! T_1$, then system $1$ is always schedulable. 

For system $2$, we first examine the schedulability of one complete period $[t_{w_0},t_{w_0}\!+\!T_2]$ of system $2$, where $t_{w_0}$ is the time when system $2$ starts a new request, i.e., $t_{w_0}\!=\!\alpha_2\!+\!(k\!-\!1)T_2$ for some $k \!\ge\! 1$. For this specific time interval, $d_2(t_{w_0})\!=\!T_2$, $r_2(t_{w_0})\!=\!C_2$, $N_c\!=\!0$ and $\tau_{{\rm e},2}\!=\!0$. Therefore, ${\rm OP}_2(t_{w_0}, t_{w_0}\!+\!T_2, d_2(t_{w_0}), r_2(t_{w_0}))\!=\!C_2$, which is a constant. 

For system $1$, since it has the highest priority, the time-occupancy for system $1$ can be represented as ${\rm OP}_1(t_{w_0}, t_{w_0}\!+\!T_2, d_1(t_{w_0}))$ which does not depend on $r_1(t_{w_0})$ based on (\ref{Remaining_Time_1}). 
\begin{lemma}
The time-occupancy of system $1$ can be computed as
\begin{align}\label{OP_1}
\begin{array}{c}
{\rm OP}_1(t_{w_0}, t_{w_0}\!+\!T_2, d_1(t_{w_0}))\!=\!\max\left\{0,d_1(t_1)\!-\!(T_1\!-\!C_1)\right\}\!+\!C_1\!\left\lfloor\!\frac{T_2\!-\!d_1(t_{w_0})}{T_1}\!\right\rfloor\!\!+\!\tau_{{\rm e},1}\\
=\!\left\{\!\!
\begin{array}{l}
C_1\!\left\lfloor\!\frac{T_2\!-\!d_1(t_{w_0})}{T_1}\!\right\rfloor\!\!+\!\min\!\left(\!T_1\!\left\{ \frac{\!T_2\!-\!d_1(t_{w_0})}{T_1}\!\right\},C_1\!\right)\!, \text{if } 0\!<\! d_1(t_{w_0})\!\le\! T_1\!-\!C_1\\
d_1(t_{w_0})\!-\!(T_1\!-\!C_1)\!+\!C_1\!\left\lfloor\!\frac{T_2\!-\!d_1(t_{w_0})}{T_1}\!\right\rfloor\!\!+\!\min\!\left(T_1\left\{\! \frac{T_2\!-\!d_1(t_{w_0})}{T_1}\!\right\},C_1\!\right)\!,\\
\text{if } T_1\!-\!C_1< d_1(t_{w_0})\!\le\! T_1
\end{array}
\right.
\end{array}
\end{align}
\end{lemma}
\begin{figure}[t]
\centering
\includegraphics[width=4.4in]{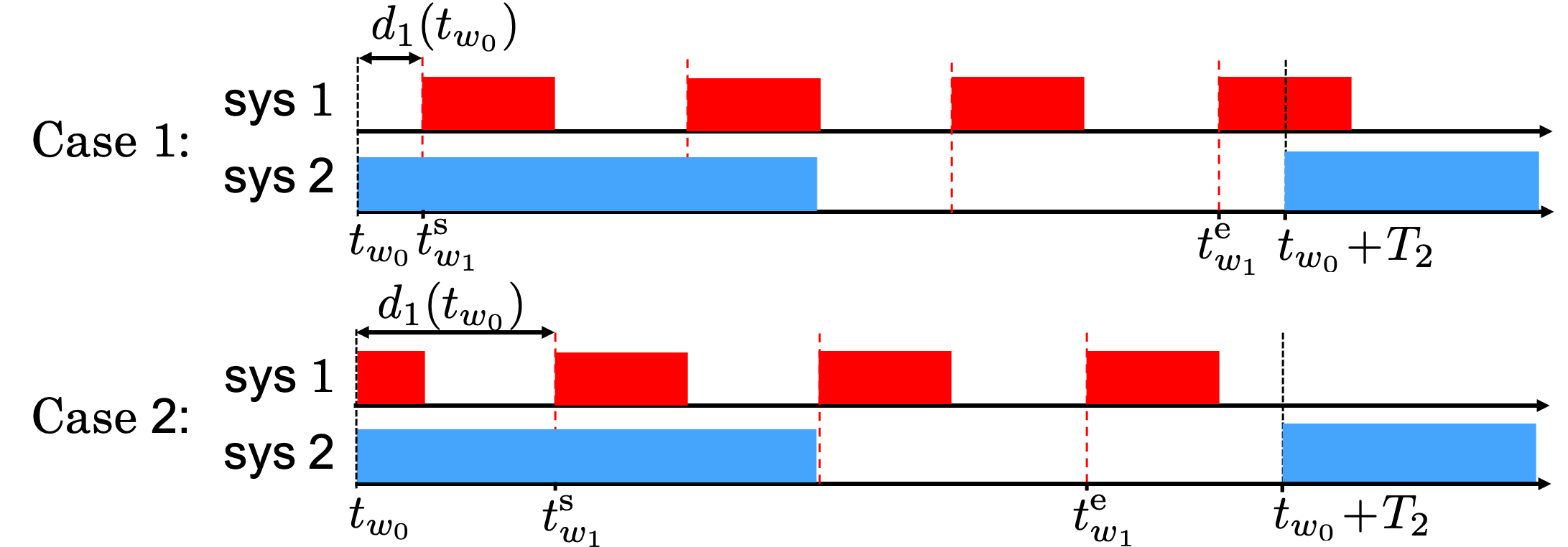}
\caption{Cases of scheduling two systems where system $1$ has higher priority. Red rectangles represent the time occupied by system $1$ and blue rectangles represent the time occupied by system $2$.}
\label{Chapter_Lemma1}\vspace{-1em}
\end{figure}
\textbf{Proof. }Due to the maximal operator in $\max\left\{0,d_1(t_1)\!-\!(T_1\!-\!C_1)\right\}$, formula (\ref{OP_1}) considers two cases depending on the values of the deadline variable $d_1(t_{w_0})$, which is illustrated by Figure \ref{Chapter_Lemma1}.\\
Case $1$: if $0\!<\! d_1(t_{w_0})\!\le \!T_1\!-\!C_1$, then $\max\left\{0,d_1(t_{w_0})\!-\!(T_1\!-\!C_1)\right\}\!=\!0$, which means that the task from system $1$ has finished using the shared resource within the first incomplete period of system $1$, i.e., $[t_{w_0},t_{w_0}\!+\!d_1(t_{w_0}))$. For simplicity, we denote $t_{w_0}\!+\!d_1(t_{w_0})$ as $t_{w_1}^{\rm s}$. We only need to compute the time occupied by system $1$ within $[t_{w_1}^{\rm s},t_{w_0}\!+\!T_2]$. We define $\lfloor\cdot\rfloor$ to be the rounding down operator, i.e. $\lfloor x\rfloor$ is the largest integer that is less than or equal to $x$. Then the number of complete periods of system $1$ within the time interval $[t_{w_1}^{\rm s},t_{w_0}\!+\!T_2]$ can be computed as 
$$N_c=\left\lfloor\frac{t_{w_0}\!+\!T_2\!-\!t_{w_1}^{\rm s}}{T_1}\right\rfloor\!=\!\left\lfloor\frac{t_{w_0}\!+\!T_2\!-\![t_{w_0}\!+\!d_1(t_{w_0})]}{T_1}\right\rfloor\!=\!\left\lfloor\frac{T_2\!-\!d_1(t_{w_0})}{T_1}\right\rfloor.$$
The ending time of the last complete period of system $1$ within the time interval $[t_{w_1}^{\rm s},t_{w_0}\!+\!T_2]$ is $t_{w_1}^{\rm s}\!+\!T_1\!\left\lfloor\frac{T_2\!-\!d_1(t_{w_0})}{T_1}\right\rfloor$, which is denoted as $t_{w_1}^{\rm e}$ for simplicity. 

Then within the last time interval, which is $\left[t_{w_1}^{\rm e},t_{w_0}\!+\!T_2\right]$, the maximal amount of time which can be occupied by system $1$ is $C_1$ since this time interval is an incomplete period of system $1$. However, the time occupied by system $1$ may be less than $C_1$ if the total duration of $\left[t_{w_1}^{\rm e},t_{w_0}\!+\!T_2\right]$ is less than $C_1$, which is illustrated by Case $1$ in Figure \ref{Chapter_Lemma1}. Therefore, we need to compare the value of the duration and $C_1$ and take the smaller value as the time that can be occupied by system $1$. The duration of $\left[t_{w_1}^{\rm e},t_{w_0}\!+\!T_2\right]$ is $\left(t_{w_0}\!+\!T_2\right)\!-\!t_{w_1}^{\rm e}\!=\!\left(t_{w_0}\!+\!T_2\right)\!-\!\left(t_{w_0}\!+\!d_1(t_{w_0})\!+\!T_1\!\left\lfloor\frac{T_2-d_1(t_{w_0})}{T_1}\right\rfloor\right)\!=\!T_2\!-\!d_1(t_{w_0})\!-\!T_1\!\left\lfloor\frac{T_2-d_1(t_{w_0})}{T_1}\right\rfloor\!=\!T_1\!\left(\frac{T_2-d_1(t_{w_0})}{T_1}\right)\!-\!T_1\!\left\lfloor\frac{T_2-d_1(t_{w_0})}{T_1}\right\rfloor\!=\!T_1\!\left\{ \frac{T_2-d_1(t_{w_0})}{T_1}\right\}$ where the operator $\{\cdot\}$ takes the fractional part of a real number, i.e., $\{ x\}\!=\!x\!-\!\lfloor x\rfloor$.
Therefore, 

$$\tau_{{\rm e},1}\!=\!\min\!\left(T_1\!\left\{ \frac{T_2\!-\!d_1(t_{w_0})}{T_1}\right\},C_1\right).$$

In summary, for Case $1$, the time occupied by system $1$ within the time interval $[t_{w_0},t_{w_0}+T_2]$ can be computed as ${\rm OP}_1(t_{w_0}, t_{w_0}\!+\!T_2, d_1(t_{w_0}))\!=\!0+C_1\!\left\lfloor\frac{T_2-d_1(t_{w_0})}{T_1}\right\rfloor\!+\!\min\!\left(T_1\!\left\{ \frac{T_2\!-\!d_1(t_{w_0})}{T_1}\right\},C_1\right)$.

Case $2$: if $T_1\!-\!C_1\!<\! d_1(t_{w_0})\!\le\! T_1$, then $\max\left\{0,d_1(t_{w_0})\!-\!(T_1\!-\!C_1)\right\}\!=\!d_1(t_{w_0})\!-\!(T_1\!-\!C_1)$. The time $\tau_{{\rm e},1}$ and the number of complete periods $N_c$ are the same formula as Case $1$. Therefore, we have ${\rm OP}_1(t_{w_0}, t_{w_0}\!+\!T_2, d_1(t_{w_0}))\!=\!d_1(t_{w_0})\!-\!(T_1\!-\!C_1)\!+\!C_1\!\left\lfloor\!\frac{T_2\!-\!d_1(t_{w_0})}{T_1}\right\rfloor\!+\!\min\!\left(\!T_1\!\left\{\! \frac{T_2\!-\!d_1(t_{w_0})}{T_1}\!\right\}\!,C_1\!\right)$.

Combining the above two cases, we have the formula (\ref{OP_1}).\hfill $\square$

\begin{lemma}\label{S_lemma_1}
A task from system $2$ is schedulable within $[t_{w_0},t_{w_0}\!+\!T_2]$ if ${\rm OP}_1(t_{w_0}, t_{w_0}\!+\!T_2, d_1(t_{w_0}))\!+\!C_2\!\le\! T_2$.
\end{lemma}
\textbf{Proof. }According to the significant moment computation formula (\ref{equation:preemptive}), the significant moments within $[t_{w_0}, t_{w_0}\!+\!T_2]$ are $t_{w_0}$, $t_{w_0}\!+\!d_1(t_{w_0})$, $\dots$, $t_{w_0}\!+\!d_1(t_{w_0})\!+\!T_1\lfloor (T_2-d_1(t_{w_0}))/T_1\rfloor$ and $t_{w_0}\!+\!T_2$. We can check the instantaneous scehdulability condition at each significant moments. 

At the last significant moment $t_{w_0}\!+\!T_2$, we have $d_2((t_{w_0}\!+\!T_2)^-)\!=\!0$. The remaining time $r_2((t_{w_0}\!+\!T_2)^-)\!=\!\max\{0,C_2\!-\![T_2\!-\!{\rm OP}_1(t_{w_0}, t_{w_0}\!+\!T_2, d_1(t_{w_0}))]\}$. Since ${\rm OP}_1(t_{w_0}, t_{w_0}\!+\!T_2, d_1(t_{w_0}))\!+\!C_2\!\le\! T_2$, we have $C_2\!-\![T_2\!-\!{\rm OP}_1(t_{w_0}, t_{w_0}\!+\!T_2, d_1(t_{w_0}))]\! \le\! 0$. Therefore, $r_2((t_{w_0}\!+\!T_2)^-)\!=\!0\!=\! d_2((t_{w_0}\!+\!T_2)^-)$. The instantaneous scehdulability condition $r_2(t^-)\!\le\! d_2(t^-)$ is satisfied at the last significant moment $t_{w_0}\!+\!T_2$.

Since $[t_{w_0}, t_{w_0}\!+\!T_2]$ is one period of system 2, the timing states of system $2$ do not have jumps within $(t_{w_0}, t_{w_0}\!+\!T_2)$. Therefore, for any significant moment $t_w \!\in\! (t_{w_0}, t_{w_0}\!+\!T_2)$, since $\dot{d}_2(t)\!=\!-\!1$, we have $d_2(t_w^-)\!=\!d_2((t_{w_0}\!+\!T_2)^-)\!+\!(t_{w_0}\!+\!T_2\!-\!t_w)$. For the remaining time variable, because $\dot{r}_2(t)\!=\!-\!1$ or $0$, we have the inequality that $r_2(t_w^-)\!\le\!r_2((t_{w_0}\!+\!T_2)^-)\!+\!(t_{w_0}\!+\!T_2\!-\!t_w)$. Because $r_2((t_{w_0}\!+\!T_2)^-)\!\le\! d_2( (t_{w_0}\!+\!T_2)^-)$, we can conclude
$$r_2(t_w^-)\!\le\!r_2((t_{w_0}\!+\!T_2)^-)\!+\!(t_{w_0}\!+\!T_2\!-\!t_w)\!\le\!d_2((t_{w_0}\!+\!T_2)^-)\!+\!(t_{w_0}\!+\!T_2\!-\!t_w)\!=\!d_2(t_w^-).$$
And for the first significant moment, it is trivial that $r_2(t_{w_0})\!=\!C_2\!\le\!T_2\!=\!d_2(t_{w_0})$. 

According to \textbf{Corollary \ref{ins_schedulability}}, if tasks from system $2$ satisfy $r_2(t_w^-)\!\le\! d_2(t_w^-)$ at any significant moment $t_w\!\in\![t_{w_0}, t_{w_0}\!+\!T_2]$, then tasks from system $2$ are schedulable at any time within $[t_{w_0}, t_{w_0}\!+\!T_2]$. \hfill $\square$

To extend \textbf{Lemma \ref{S_lemma_1}} to an infinite time interval, we need to find the worst case. Since $C_2$ and $T_2$ are constant, the worst case occurs when the time-occupancy ${\rm OP}_1$ is maximized.
\begin{theorem}\label{lemma2}
The maximal value of ${\rm OP}_1(t_{w_0}, t_{w_0}\!+\!T_2, d_1(t_{w_0}))$, denoted as ${\rm OP}_{1,{\rm max}}$, can be obtained when $d_1^*(t_{w_0})\!=\!T_1$ and
\begin{align}\label{Theorem_1_OP1_formula}
{\rm OP}_{1,{\rm max}}\!=\!C_1\!\left\lfloor\frac{T_2}{T_1}\right\rfloor\!+\!\min\!\left(T_1\!\left\{ \frac{T_2}{T_1}\right\},C_1\right).    
\end{align}
\end{theorem}
\textbf{Proof. }The proof is in Appendix \ref{Proof_theorem_1}.\hfill $\square$
\begin{Remark}
Here $d_1^*(t_{w_0})\!=\!T_1$ while $d_2(t_{w_0})=T_2$ by the definition of $t_{w_0}$. Therefore, the worst-case moment is when both systems $1$ and $2$ start a task at the same time, which is actually the same worst-case moment as the critical time instant introduced in \cite{liu1973scheduling}. Such worst-case moment can be easily achieved, for example, by letting  systems $1$ and $2$ start the first task request at the same time. \textbf{Theorem \ref{lemma2}} shows the great advantage of using the significant moment analysis because the worst-case deadline and moment can be obtained and justified by derivations from  mathematical equations. 
\end{Remark}
Here is a new insight discovered using our method. Based on the value of timing characteristics, the critical time instant may not be the only worst-case moment.
\begin{corollary}\label{Coro_worst_case_moments}
If the timing characteristics satisfy $T_2\!=\!MT_1\!+\!C_1$ where $M\!\ge\!1$ is an integer, then $d_1^*(t_{w_0})\!=\!T_1$ is the only worst-case deadline such ${\rm OP}_{1,{\rm max}}$ can be obtained. If $MT_1\!<\!T_2\!<\!MT_1\!+\!C_1$, then the maximal value ${\rm OP}_{1,{\rm max}}$ can also obtained when $T_1\!-\!C_1\!+\!T_1\!\left\{ \frac{T_2}{T_1}\right\}\!\le\! d_1^*(t_{w_0})\!\le\!T_1$. If $MT_1\!+\!C_1\!<\!T_2\!<\!(M\!+\!1)T_1$, then the maximal value can also be obtained when $0\!<\! d_1^*(t_{w_0})\!<\! T_1\!\left\{ \frac{T_2}{T_1}\right\}\!-\!C_1$. If $T_2\!=\!MT_1$, then ${\rm OP}_{1,{\rm max}}$ can be achieved for any $d_1(t_{w_0})$ satisfying $0\!<\!d_1(t_{w_0})\!\le\!T_1$.
\end{corollary}
\textbf{Proof. }The proof is in Appendix \ref{Proof_corollary_worst_moment}.\hfill $\square$

\textbf{Corollary \ref{Coro_worst_case_moments}} shows that our analysis based on significant moments can find all worst case deadlines rather than the one obtained at the critical time instant.

Based on the \textbf{Theorem \ref{lemma2}}, we can have
\begin{corollary}
Tasks from system $2$ are schedulable at any time $t$ satisfying $t \!\ge\! \alpha_2$, if $C_1\!\left\lfloor T_2/T_1\right\rfloor\!+\!\min\!\left(T_1\left\{ T_2/T_1\right\},C_1\right)\!+\!C_2\! \le\! T_2$, where $\alpha_2$ is the time instant when system $2$ makes the first request to use the shared resource.
\end{corollary}
\textbf{Proof. } If a task from system $2$ is schedulable at the worst-case moments, then it is always schedulable for one period. Since system $2$ are periodic, the worst cases are the same for each period. Therefore, the schedulability is satisfied for any period of system $2$. \hfill $\square$
\subsubsection{System $2$ has higher priority}
The other priority assignment is that system $2$ has higher priority than system $1$, i.e. $p_2(t)\!<\!p_1(t)$ for all $t$. It is trivial that if $C_2\!<\! T_2$, then system $2$ is always schedulable.

For system $1$, we consider the schedulability in one complete period $[t_{w_0}',t_{w_0}'\!+\!T_1]$ of system $1$ where $t_{w_0}'$ is the starting of a new task of system $1$, i.e., $t_{w_0}'\!=\!\alpha_1\!+\!(k\!-\!1)T_1$ for some $k \!\ge\! 1$. For system $1$, we have $d_1(t_{w_0}')\!=\!T_1$, $r_1(t_{w_0}')\!=\!C_1$, $N_c\!=\!0$ and $\tau_{{\rm e},1}\!=\!0$. Therefore, ${\rm OP}_1(t_{w_0}', t_{w_0}'\!+\!T_1, d_1(t_{w_0}'), r_1(t_{w_0}'))\!=\!C_1$. 

\begin{figure}[t]
\centering
\includegraphics[width=4.4in]{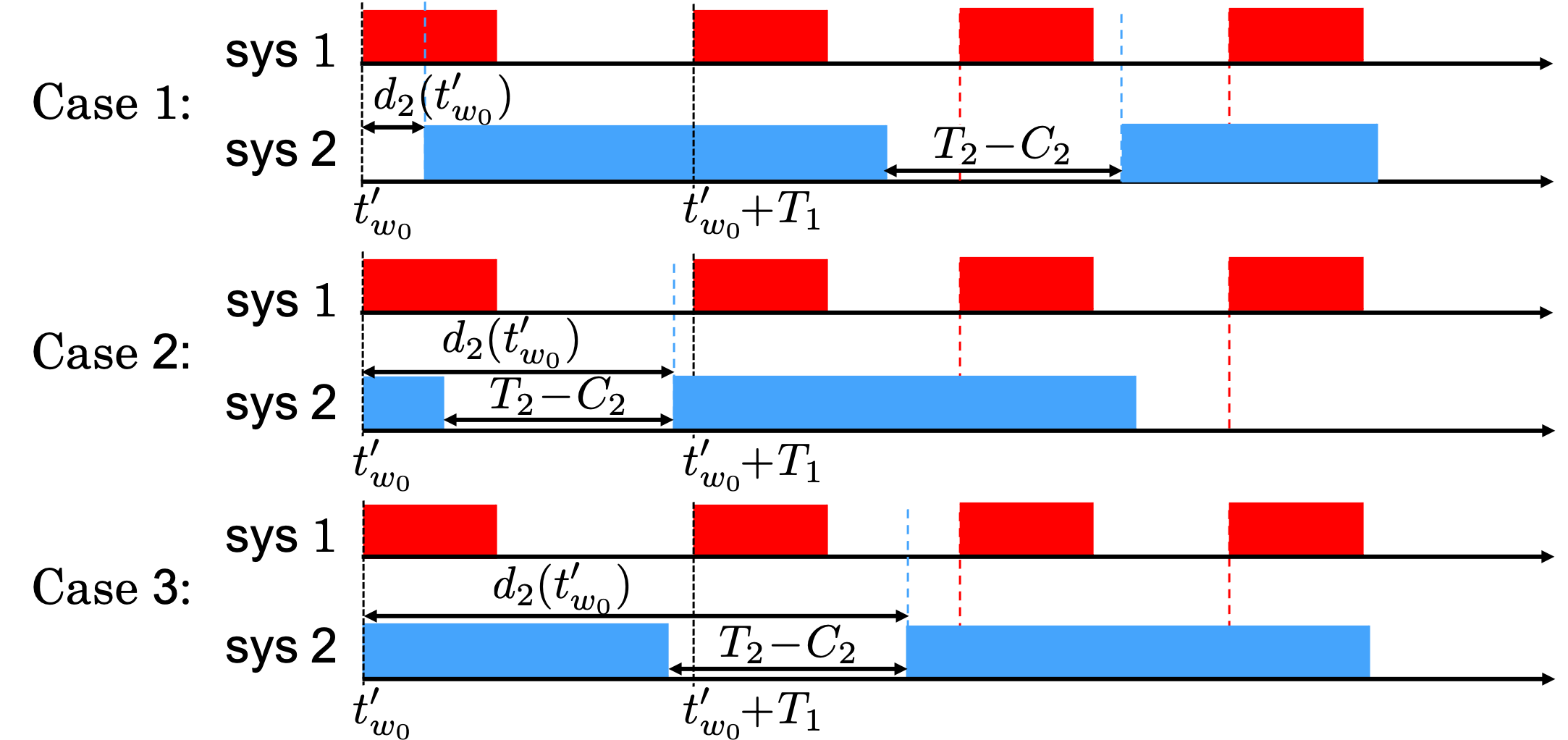}
\caption{Cases of scheduling two systems where system $2$ has higher priority. Red rectangles represent the time occupied by system $1$ and blue rectangles represent the time occupied by system $2$.}
\label{Chapter_Lemma_system_2}
\end{figure}

Since the time interval we consider is one period of system $1$ and $T_1\!<\!T_2$, there is only one incomplete period of system $2$ within $[t_{w_0}',t_{w_0}'\!+\!T_1]$, i.e., $N_c\!=\!0$. Since system $2$ has the highest priority, the time-occupancy ${\rm OP}_2(t_{w_0}', t_{w_0}'\!+\!T_1, d_2(t_{w_0}'))$ which does not depends on $r_2(t_{w_0}')$. 
\begin{lemma}
The time-occupancy of system $2$ can be computed as
\begin{align}\label{OP_2}
\begin{array}{c}
{\rm OP}_2(t_{w_0}', t_{w_0}'\!+\!T_1, d_2(t_{w_0}'))\!=\!
\left\{
\begin{array}{l}
\min\left(T_1\!-\!d_2(t_{w_0}'),C_2\right),\,\text{if } 0\!<\! d_2(t_{w_0}')\!\le\! T_2\!-\!C_2\\
T_1\!-\!T_2\!+\!C_2,\text{if } T_2\!-\!C_2\!<\! d_2(t_{w_0}')\!\le\! T_1\\
d_2(t_{w_0}')\!-\!(T_2\!-\!C_2),\text{if } T_1\!<\!d_2(t_{w_0}')\!\le\! T_2
\end{array}
\right.
\end{array}
\end{align}
\end{lemma}
\textbf{Proof. }The formula to compute ${\rm OP}_2(t_{w_0}', t_{w_0}'\!+\!T_1, d_2(t_{w_0}'))$ needs to consider three cases, which is illustrated by Figure \ref{Chapter_Lemma_system_2}.
\\
Case $1$: if $0\!<\! d_2(t_{w_0}')\!\le\! T_2\!-\!C_2$, then $\max\left\{0,d_2(t_{w_0}')\!-\!(T_2\!-\!C_2)\right\}\!=\!0$. A new task from system $2$ is generated at time $t_{w_0}'\!+\!d_2(t_{w_0}')$. For $\tau_{{\rm e},2}$, if the occupation of this task ends earlier than or at $t_{w_0}'\!+\!T_1$, then $\tau_{{\rm e},2}\!=\!C_2$. If the occupation of this task ends later than $t_{w_0}'\!+\!T_1$, as illustrated by Case $1$ in Figure \ref{Chapter_Lemma1}, then $\tau_{{\rm e},2}\!=\!T_1\!-\!d_2(t_{w_0}')$. Therefore, we have ${\rm OP}_2(t_{w_0}', t_{w_0}'\!+\!T_1, d_2(t_{w_0}'))\!=\!\tau_{{\rm e},2}\!=\min\!\left(T_1-d_2(t_{w_0}'),C_2\right)$. \\
Case $2$: if $T_2-C_2\!<\! d_2(t_{w_0}')\!\le\! T_1$, then $\max\left\{0,d_2(t_{w_0}')\!-\!(T_2\!-\!C_2)\right\}\!=\!d_2(t_{w_0}')\!-\!(T_2\!-\!C_2)$. And since $d_2(t_{w_0}')\!\le\! T_1$, the starting time of next task $t_{w_0}'\!+\!d_2(t_{w_0}')$ from system $2$ is within the interval $[t_{w_0}',t_{w_0}'\!+\!T_1]$. We have $\tau_{{\rm e},2}\!=\!T_1\!-\!d_2(t_{w_0}')$. Therefore, ${\rm OP}_2(t_{w_0}', t_{w_0}'\!+\!T_1, d_2(t_{w_0}'))\!=\!d_2(t_{w_0}')\!-\!(T_2\!-\!C_2)\!+\!T_1\!-\!d_2(t_{w_0}')\!=\!T_1\!-\!T_2\!+\!C_2$. \\
Case $3$: if $T_1\!<\!d_2(t_{w_0}')\!\le\! T_2$, then part of the task from system $2$ is executing at time $t_{w_0}'$ and the generation time $t_{w_0}'\!+\!d_2(t_{w_0}')$ of next task from system $2$ is greater than $t_{w_0}'\!+\!T_1$. Therefore, $\tau_{{\rm e},2}\!=\!0$. The time-occupancy of the task from system $2$ is ${\rm OP}_2(t_{w_0}', t_{w_0}'\!+\!T_1, d_2(t_{w_0}'))\!=\!d_2(t_{w_0}')\!-\!(T_2\!-\!C_2)$.\hfill $\square$

\begin{lemma}\label{S_lemma_2}
A task from system $1$ is schedulable within $[t_{w_0}',t_{w_0}'\!+\!T_1]$ if ${\rm OP}_2(t_{w_0}', t_{w_0}'\!+\!T_1, d_2(t_{w_0}'))\!+\!C_1\!\le\! T_1$.
\end{lemma}
\textbf{Proof.} Similar as the proof for \textbf{Lemma \ref{S_lemma_1}}. At the significant moment $t_{w_0}'\!+\!T_1$, we have $d_1((t_{w_0}'\!+\!T_1)^-)\!=\!0$. The remaining time $r_1((t_{w_0}'\!+\!T_1)^-)\!=\!\max\{0,C_1\!-\![T_1\!-\!{\rm OP}_2(t_{w_0}', t_{w_0}'\!+\!T_1, d_2(t_{w_0}'))]\}$. Since ${\rm OP}_2(t_{w_0}', t_{w_0}'\!+\!T_1, d_2(t_{w_0}'))\!+\!C_1\!\le\! T_1$, we have $C_1\!-\![T_1\!-\!{\rm OP}_2(t_{w_0}', t_{w_0}'\!+\!T_1, d_2(t_{w_0}'))]\! \le\! 0$. Therefore, $r_1((t_{w_0}'\!+\!T_1)^-)\!\le\! d_1((t_{w_0}'\!+\!T_1)^-)$ is satisfied. With $\dot{d}_1(t)\!=\!-\!1$ and $\dot{r}_1(t)\!=\!-\!1$ or $0$, we have  $r_1(t_{w}^-)\!\le\! d_1(t_{w}^-)$ for any $t_w\!\in\![t_{w_0}',t_{w_0}'\!+\!T_1]$. \hfill $\square$

\begin{theorem}\label{lemma_max}
The maximal value of ${\rm OP}_2(t_{w_0}', t_{w_0}'\!+\!T_1, d_2(t_{w_0}'))$, denoted as ${\rm OP}_{2,{\rm max}}$, equals to $C_2$. If the timing characteristic satisfies $T_1\!\le\!C_2$, then this maximal value is obtained only when $d_2^*(t_{w_0}')\!=\!T_2$. If $T_1\!>\!C_2$, then the maximal value can be obtained when $0\!<\! d_2^*(t_{w_0}')\! \le\! T_1\!-\!C_2$ or $d_2^*(t_{w_0}')\!=\!T_2$.    
\end{theorem}
\textbf{Proof. }The proof is in Appendix \ref{Proof_theorem_2}.\hfill $\square$

\begin{Remark}
Here again, we can show that significant moment analysis has found all worst case moments other than the critical instant. If the timing characteristic satisfies $T_1\!>\!C_2$, then ${\rm OP}_2$ reaches its maximal value $C_2$ when $d_2^*(t_{w_0}')$ is within the range $(0,T_1\!-\!C_2]$ or $d_2^*(t_{w_0}')\!=\!T_2$ while $d_1(t_{w_0}')\!=\!T_1$. The critical time instant when $d_1(t_{w_0}')\!=\!T_1$ and $d_2(t_{w_0}')\!=\!T_2$ is only one solution which makes ${\rm OP}_2$ reaches it’s maximal value.
\end{Remark}
Based on the \textbf{Theorem \ref{lemma_max}}, we can have
\begin{corollary}
Tasks from system $1$ are schedulable at any time $t \!\ge\! \alpha_1$, if $C_1\!+\!C_2\! \le\! T_1$, where $\alpha_1$ is the time instant when system $1$ makes the first request to use the shared resource.
\end{corollary}
\textbf{Proof. } Since the task from system $1$ is schedulable at the worst-case moments of one period and both systems $1$ and $2$ are periodic, then system $1$ is always schedulable for any time $t \!\ge\! \alpha_1$.  \hfill $\square$

\subsubsection{System 1 should have higher priority}
Then we will prove a better priority assignment is that system $1$ has higher priority than system $2$ in the sense of schedulability, which is the priority assignment under RMS.
\begin{theorem}\label{theorem_RMS}
If system $2$ is assigned with higher priority and all the tasks are schedulable at any time, then tasks must also be schedulable if system $1$ is assigned with higher priority than system $2$.        
\end{theorem}
\textbf{Proof.} We will show that $C_1\!+\!C_2 \!\le\! T_1$ implies $C_1\!\left\lfloor T_2/T_1\right\rfloor\!+\!\min\left(T_1\left\{ T_2/T_1\right\},C_1\right)\!+\!C_2 \!\le\! T_2$. Since $T_2\!>\! T_1$, we have $\left\lfloor T_2/T_1\right\rfloor \!\ge\! 1$ and
\begin{align}
 C_1\!\left\lfloor T_2/T_1\right\rfloor\!+\!\min\!\left(T_1\!\left\{ T_2/T_1\right\},C_1\right) &\le C_1\!\left\lfloor T_2/T_1\right\rfloor\!+\!T_1\!\left\{ T_2/T_1\right\}\!+\!C_2\nonumber\\
 &\le C_1\!\left\lfloor T_2/T_1\right\rfloor\!+\!T_1\!\left\{ T_2/T_1\right\}\!+\!C_2\!\left\lfloor T_2/T_1\right\rfloor\nonumber\\
 &\!=\!(C_1\!+\!C_2)\!\left\lfloor T_2/T_1\right\rfloor\!+\!T_1\!\left\{ T_2/T_1\right\}.
\end{align}
Because $C_1\!+\!C_2 \!\le\! T_1$, we have $(C_1\!+\!C_2)\!\left\lfloor T_2/T_1\right\rfloor\!+\!T_1\!\left\{ T_2/T_1\right\}\!\le\! T_1\!\left\lfloor T_2/T_1\right\rfloor\!+\!T_1\!\left\{ T_2/T_1\right\}\!=\!T_1\cdot T_2/T_1\!=\!T_2$. \hfill $\square$
\subsection{Schedulability for $N$ Systems}
 We now extend our results to $N\!>\!2$ systems using the results of the two systems. First, we leverage the following result from \cite{liu1973scheduling} which shows that RMS is the optimal static scheduling strategy
\begin{corollary}\label{Coro_N_Systems}
If a feasible fixed priority assignment exists for some task set, the RMS priority assignment is feasible for that task set.        
\end{corollary}
\textbf{Proof. }The proof for this corollary can be found in Theorem $2$ in \cite{liu1973scheduling}.\hfill $\square$

Based on \textbf{Corollary \ref{Coro_N_Systems}}, RMS is the optimal static scheduling methods. In the rest of this subsection, we will directly use the priority assignments under RMS, i.e., if there are systems $j$ and $i$ with indices $j\!<\!i$, then $p_j\!<\!p_i$ because $T_j\!<\!T_i$.

The worst-case schedulability condition for systems $1$ and $2$ are the same as the two system case. Here, we will consider the schedulability for system $i\!>\!2$ in one period $[t_{w_0}, t_{w_0}\!+\!T_i]$ where $t_{w_0}$ is the significant moment when system $i$ starts a new task and $T_i$ is the period of system $i$. For this time interval, $d_i(t_{w_0})\!=\!T_i$, $r_i(t_{w_0})\!=\!C_i$, $N_c\!=\!0$ and $\tau_{{\rm e},i}\!=\!0$. Therefore, ${\rm OP}_i(t_{w_0}, t_{w_0}\!+\!T_i, d_i(t_{w_0}), r_i(t_{w_0}))\!=\!C_i$, which is a constant. 


For system $1$, since it has the highest priority under RMS, the time-occupancy does not depend on $r_1(t_{w_0})$ and cannot be interrupted by other systems. The time-occupancy ${\rm OP}_1(t_{w_0}, t_{w_0}\!+\!T_i, d_1(t_{w_0}))$ can be computed similarly as (\ref{OP_1}) with system $i$ replacing system $2$
\begin{align}\label{OP_N_systems}
\begin{array}{c}
\hspace{-18em}{\rm OP}_1(t_{w_0}, t_{w_0}\!+\!T_i, d_1(t_{w_0}))\!=\!\\
\!\left\{\!\!
\begin{array}{l}
C_1\!\left\lfloor\!\frac{T_i\!-\!d_1(t_{w_0})}{T_1}\!\right\rfloor\!\!+\!\min\!\left(\!T_1\!\left\{ \frac{\!T_i\!-\!d_1(t_{w_0})}{T_1}\!\right\},C_1\!\right)\!, \text{if } 0\!<\! d_1(t_{w_0})\!\le\! T_1\!-\!C_1\\
d_1(t_{w_0})\!-\!(T_1\!-\!C_1)\!+\!C_1\!\left\lfloor\!\frac{T_i\!-\!d_1(t_{w_0})}{T_1}\!\right\rfloor\!\!+\!\min\!\left(T_1\left\{\! \frac{T_i\!-\!d_1(t_{w_0})}{T_1}\!\right\},C_1\!\right)\!,\\
\text{if } T_1\!-\!C_1< d_1(t_{w_0})\!\le\! T_1
\end{array}
\right.
\end{array}
\end{align}

\begin{figure}[t]
\centering
\includegraphics[width=4.4in]{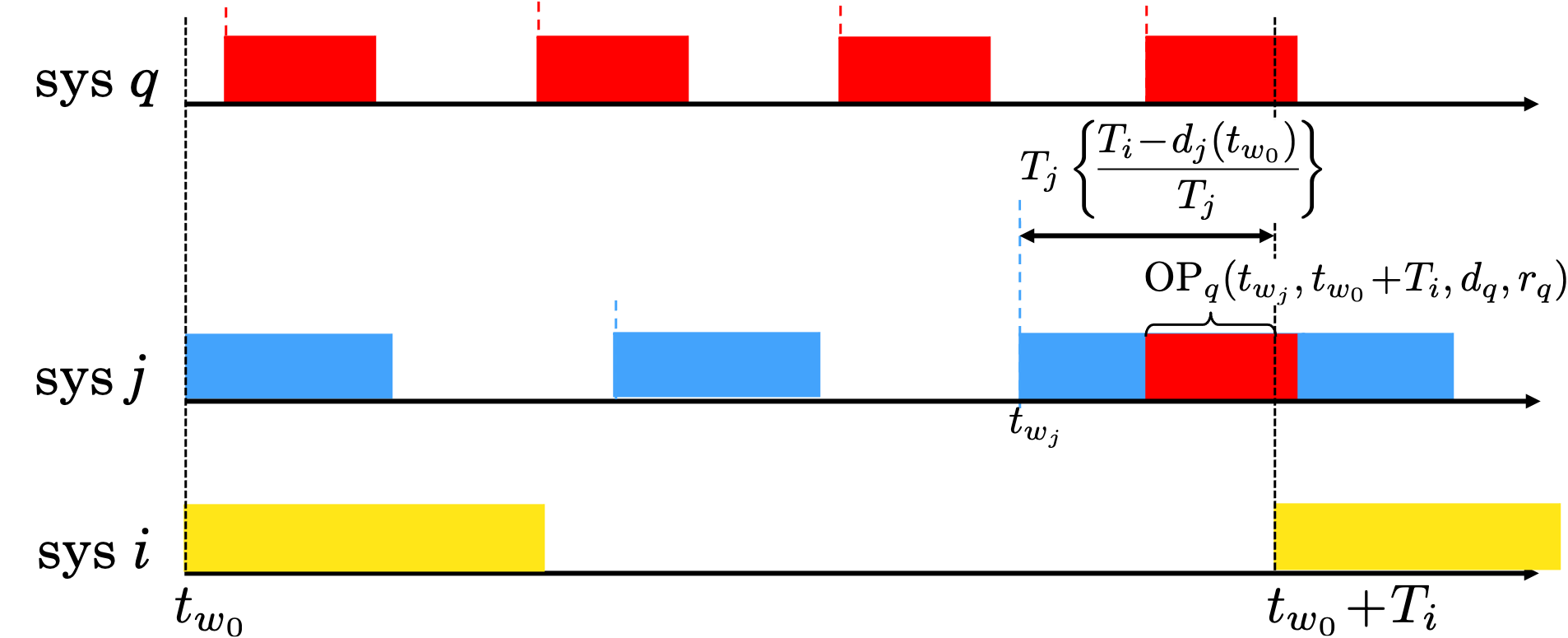}
\caption{Illustration time-occupancies of $N$ systems.}
\label{therorem_figure_Occupation_time_n_systems}
\end{figure}

For system $j\!=\!2,...,i\!-\!1$, its resource occupation may be interrupted by higher prioritized system. In each complete period of system $j$ within $[t_{w_0}, t_{w_0}\!+\!T_i]$, the time-occupancy of system $j$ should still be $C_j$ even though the occupation be interrupted by higher prioritized system. Otherwise, system $j$ is unschedulable. Therefore, the first two terms in formula (\ref{formula_general}) are the same when computing ${\rm OP}_j(t_{w_0}, t_{w_0}\!+\!T_i, d_j(t_{w_0}), r_j(t_{w_0}))$. The difference occurs when considering $\tau_{{\rm e},j}$ in the last incomplete period of system $j$, which is $\left[t_{w_0}\!+\!d_j(t_{w_0})\!+\!T_j\!\left\lfloor\frac{T_i\!-\!d_j(t_{w_0})}{T_j}\right\rfloor, t_{w_0}\!+\!T_i\right]$. We use the notation $t_{w_j}$ to represent the starting time of the last incomplete period $t_{w_0}\!+\!d_j(t_{w_0})\!+\!T_j\!\left\lfloor\frac{T_i\!-\!d_j(t_{w_0})}{T_j}\right\rfloor$ for simplicity. The time occupied by a higher prioritized system $q$ where $q\!<\!j$ can be denoted as ${\rm OP}_q\left(t_{w_j}, t_{w_0}\!+\!T_i, d_q(t_{w_j}), r_q(t_{w_j})\right)$ based on \textbf{Definition \ref{Def_occupancy}}. As illustrated in Figure \ref{therorem_figure_Occupation_time_n_systems}, the time interval which can be used by system $j$ is
$T_j\left\{\! \frac{T_i\!-\!d_j(t_{w_0})}{T_j}\!\right\}-\sum_{q=1}^{j-1}{\rm OP}_q\left(t_{w_j}, t_{w_0}\!+\!T_i, d_q(t_{w_j}), r_q(t_{w_j})\right)$ rather than $T_j\left\{\! \frac{T_i\!-\!d_j(t_{w_0})}{T_j}\!\right\}$. If it is less than $C_j$, then this whole time interval will be occupied by system $j$. If it is greater than or equal to $C_j$, then the time occupied by system $j$ is $C_j$. Therefore, the time occupied by system $j$ in the last incomplete period can be computed as 
\begin{align}\label{OP_j_last}
\tau_{{\rm e},j}\!=\!\min\!\left(\! T_j\left\{\! \frac{T_i\!-\!d_j(t_{w_0})}{T_j}\!\right\}\!-\!\sum_{q=1}^{j-1}{\rm OP}_q\left(t_{w_j}, t_{w_0}\!+\!T_i, d_q(t_{w_j}), r_q(t_{w_j})\right), C_j\!\right).
\end{align}
The time-occupancy ${\rm OP}_j(t_{w_0}, t_{w_0}\!+\!T_i, d_j(t_{w_0}), r_j(t_{w_0}))$ can be computed as
\begin{align}\label{OP_j_systems}
{\rm OP}_j(t_{w_0}, t_{w_0}\!+\!T_i, d_j(t_{w_0}), r_j(t_{w_0}))\!=\!r_j(t_{w_0})\!+\!C_j\!\left\lfloor\!\frac{T_i\!-\!d_j(t_{w_0})}{T_j}\!\right\rfloor\!+\!\tau_{{\rm e},j}.
\end{align}
Based on the formula (\ref{OP_j_systems}), we have the following lemma
\begin{lemma}\label{lemma_principle}
If the timing state variables $d_j(t_{w_0})$ and $r_j(t_{w_0})$ of system $j$ is unchanged, then the time-occupancy ${\rm OP}_j(t_{w_0}, t_{w_0}\!+\!T_i, d_j(t_{w_0}), r_j(t_{w_0}))$ will only be decreased if there is an increase in the time-occupancy of at least one system $q$ which has higher priority than system $j$.
\end{lemma}
\textbf{Proof.} Since $d_j(t_{w_0})$ and $r_j(t_{w_0})$ are unchanged, the first term $r_j(t_{w_0})$ and second term $C_j\!\left\lfloor\!\frac{T_i\!-\!d_j(t_{w_0})}{T_j}\!\right\rfloor$ in (\ref{OP_j_systems}) remain unchanged. As for the term $\tau_{{\rm e},j}$, $T_j\left\{\! \frac{T_i\!-\!d_j(t_{w_0})}{T_j}\!\right\}$ and $C_j$ is also unchanged. Therefore, if the value of ${\rm OP}_j(t_{w_0}, t_{w_0}\!+\!T_i, d_j(t_{w_0}), r_j(t_{w_0}))$ is decreased, then all the decrements must be due to the increments in the term $\sum_{q=1}^{j-1}{\rm OP}_q\left(t_{w_j}, t_{w_0}\!+\!T_i, d_q(t_{w_j}), r_q(t_{w_j})\right)$ where at least one time-occupancy ${\rm OP}_q$ must be increased. \hfill $\square$

This lemma shows an intuitive principle of preemptive systems: if the schedule of a lower prioritized system is not changed within a finite time interval but its time-occupancy is reduced, then the reduced amount of time must all be occupied by other higher prioritized systems. 

\begin{lemma}\label{lemma_N_systems}
When $i \!\ge\! 2$, then system $i$  is schedulable within time $[t_{w_0},t_{w_0}\!+\!T_i]$ if $$\sum_{j=1}^{i-1}{\rm OP}_j(t_{w_0}, t_{w_0}\!+\!T_i, d_j(t_{w_0}), r_j(t_{w_0}))\!+\!C_i\!\le\! T_i.$$
\end{lemma}
\textbf{Proof.} Since the optimal fixed priority assignment is RMS, for any task from system $i$, the tasks from system with index $j\!<\!i$ have higher priority than $i$. For the task from system $i$ to be schedulable, the total time-occupancy of all the higher prioritized tasks should be less than or equal to $T_i\!-\!C_i$. \hfill $\square$

\begin{theorem}\label{theorem_N_system}
The maximal value of $\sum_{j=1}^{i-1}{\rm OP}_j(t_{w_0}, t_{w_0}\!+\!T_i, d_j(t_{w_0}), r_j(t_{w_0}))$ can be obtained when $d_j^*(t_{w_0})\!=\!T_j$ and $r_j^*(t_{w_0})\!=\!C_j$ where $j\!=\!1,...,i\!-\!1$.   
\end{theorem}
\textbf{Proof. }The detailed proof is in Appendix \ref{Proof_theorem_N_system}.\hfill $\square$

This theorem shows that the worst-case occurs when tasks from all systems are generated at the same time, which is the critical instant introduced in Theorem $1$ in \cite{liu1973scheduling}. 

Denoting the maximal value of $\sum_{j=1}^{i-1}{\rm OP}_j(t_{w_0}, t_{w_0}\!+\!T_i, d_j(t_{w_0}), r_j(t_{w_0}))$ as $\left(\sum_{j=1}^{i-1}{\rm OP}_{j}\right)_{\rm max}$, we have the worst-case schedulability test for tasks from system $i$ as
\begin{corollary}
System $i$ is schedulable at any time $t \!\ge\! \alpha_i$ if $\left(\sum_{j=1}^{i-1}{\rm OP}_{j}\right)_{\rm max}\!+\!C_i\! \le\! T_i$.
\end{corollary}
\textbf{Proof.} Since the optimal fixed priority assignment is RMS, for any task $i$, the tasks with index $j\!<\!i$ have higher priority than $i$. Therefore, the worst-case moment for task $i$ is the time-occupancy of the higher prioritized systems reach their maximal and $\left(\sum_{j=1}^{i-1}{\rm OP}_{j}\right)_{\rm max}\!+\!C_i$ is less than or equal to one period of system $i$. \hfill $\square$

\subsection{The least Upper Bound of Utilization}
In real-time system, the utilization factor $U\!=\!\sum_{i=1}^N (C_i/T_i)$ is used to test feasibility. Since RMS is the optimal fixed priority assignment, the utilization factor achieved by the RMS is greater than or equal to the utilization factor for any other priority assignment for that task set. Therefore, the least upper bound of the utilization factor is the infimum of the utilization factors corresponding to the RMS over all possible
task periods and run-times for the tasks. The bound is first determined for two tasks, then extended for an arbitrary number of tasks.
\begin{theorem}\label{theorem_2_bound}
For a set of two tasks with fixed priority assignment, the least upper bound on the processor utilization factor is $\overline{U}_{\rm min}\!=\!2(2^{\frac{1}{2}}\!-\!1)$.
\end{theorem}
\textbf{Proof. }The proof is in Appendix \ref{Proof_theorem_bound}.\hfill $\square$

The following results in \textbf{Corollary \ref{utilization_bound_2}} and \textbf{\ref{utilization_bound_3}} of untilization bounds have been shown in Theorem $4$ and $5$ in \cite{liu1973scheduling}. Here we recap the statements of these results.
\begin{corollary}\label{utilization_bound_2}
For a set of $N$ tasks with fixed priority order, and the restriction that the ratio between any two task periods is less than $2$, the least upper bound to the processor utilization factor is $\overline{U}_{\rm min}\!=\!N(2^{\frac{1}{N}}\!-\!1)$.
\end{corollary}
\begin{corollary}\label{utilization_bound_3}
For a set of $N$ tasks with fixed priority order, the least upper bound to the processor utilization factor is $\overline{U}_{\rm min}\!=\!N(2^{\frac{1}{N}}\!-\!1)$.
\end{corollary}

\section{Conclusion}\label{Conclusion}
In this paper, we presented a method to derive infinite-time schedulability for a set of real-time systems on a uniprocessor through mathematical derivation. And we discovered that the critical time instant is not the only worst-case moment. The schedulability result examined for preemptive and periodic tasks can reproduce the classical results.


\section{Appendix}\label{Appendix}
\subsection{Proof of Theorem \ref{lemma2}}\label{Proof_theorem_1}
Because (\ref{OP_1}) to compute ${\rm OP}_1(t_{w_0}, t_{w_0}\!+\!T_2, d_1(t_{w_0}))$ is derived in two cases, we will also discuss when ${\rm OP}_1(t_{w_0}, t_{w_0}\!+\!T_2, d_1(t_{w_0}))$ reaches maximal value in two cases.\vspace{0.5em}\\
{\bf Case $\bf 1$} is $0\!<\! d_1(t_{w_0})\!\le\! T_1\!-\!C_1$. We will show that ${\rm OP}_1(t_{w_0}, t_{w_0}\!+\!T_2, d_1(t_{w_0}))$ is a non-increasing function of $d_1(t_{w_0})$ in the range of $(0,T_1-C_1]$. Therefore, the supreme of ${\rm OP}_1(t_{w_0}, t_{w_0}\!+\!T_2, d_1(t_{w_0}))$ is obtained when $d_1(t_{w_0})\!=\!0$. Define a variable $d_1'(t_{w_0})$ which satisfies $0\!<\! d_1'(t_{w_0})\!<\!d_1(t_{w_0})\!\le\! T_1\!-\!C_1$ and its corresponding time-occupancy is denoted as ${\rm OP}_1(t_{w_0}, t_{w_0}\!+\!T_2, d_1'(t_{w_0}))$.  We will show that ${\rm OP}_1(t_{w_0}, t_{w_0}\!+\!T_2, d_1'(t_{w_0}))\!\ge\! {\rm OP}_1(t_{w_0}, t_{w_0}\!+\!T_2, d_1(t_{w_0}))$ is always satisfied by taking the difference between these two time-occupancies as $\Delta_1\!=\!{\rm OP}_1(t_{w_0}, t_{w_0}\!+\!T_2, d_1'(t_{w_0}))\!-\!{\rm OP}_1(t_{w_0}, t_{w_0}\!+\!T_2, d_1(t_{w_0}))$ and proving $\Delta_1\ge 0$. 

Plugging in the formula of time-occupancy, we have
\begin{align}\label{difference_1}
\Delta_1&\!=\!C_1\!\left(\left\lfloor\frac{T_2-d_1'(t_{w_0})}{T_1}\right\rfloor\!-\!\left\lfloor\frac{T_2-d_1(t_{w_0})}{T_1}\right\rfloor\right)\!\!+\!\min\!\left(T_1\!\left\{ \frac{T_2-d_1'(t_{w_0})}{T_1}\right\},C_1\!\right)\nonumber\\
&\!-\!\min\!\left(T_1\!\left\{ \frac{T_2-d_1(t_{w_0})}{T_1}\right\},C_1\right)
\end{align}
where $\left\lfloor\!\frac{T_2-d_1'(t_{w_0})}{T_1}\!\right\rfloor$ and $\left\{\!\frac{T_2-d_1'(t_{w_0})}{T_1}\!\right\}$ are the integer and fractional parts of $\frac{T_2-d_1'(t_{w_0})}{T_1}$, respectively. Similarly, $\left\lfloor\!\frac{T_2-d_1(t_{w_0})}{T_1}\!\right\rfloor$ and $\left\{\!\frac{T_2-d_1(t_{w_0})}{T_1}\!\right\}$ are the integer and fractional parts of $\frac{T_2-d_1(t_{w_0})}{T_1}$. 

The term $\left\lfloor\!\frac{T_2-d_1'(t_{w_0})}{T_1}\!\right\rfloor\!-\!\left\lfloor\!\frac{T_2-d_1(t_{w_0})}{T_1}\!\right\rfloor$ in above equation can only take two values, either $0$ or $1$. The reasons are as follows. First, we analyze the values of $\frac{T_2-d_1'(t_{w_0})}{T_1}$ and $\frac{T_2-d_1(t_{w_0})}{T_1}$ by taking the difference $\frac{T_2-d_1'(t_{w_0})}{T_1}-\frac{T_2-d_1(t_{w_0})}{T_1}=\frac{d_1(t_{w_0})-d_1'(t_{w_0})}{T_1}$. Then, because of the defined range $0\!<\! d_1'(t_{w_0})\!<\!d_1(t_{w_0})\!\le\! T_1\!-\!C_1$, we have $0<d_1(t_{w_0})\!-\!d_1'(t_{w_0})\!<\!T_1-C_1\!<\!T_1$, which leads to $0\!<\!\frac{d_1(t_{w_0})-d_1'(t_{w_0})}{T_1}\!<\!\frac{T_1}{T_1}\!=\!1$. Therefore, we can conclude that $0\!<\!\frac{T_2-d_1'(t_{w_0})}{T_1}\!-\!\frac{T_2-d_1(t_{w_0})}{T_1}\!<\!1$. Based on the range of the difference of these two real numbers, the integer parts of them must satisfy $0\le \left\lfloor\frac{T_2-d_1'(t_{w_0})}{T_1}\right\rfloor\!-\!\left\lfloor\frac{T_2-d_1(t_{w_0})}{T_1}\right\rfloor\le 1$. Therefore, $\left\lfloor\frac{T_2-d_1'(t_{w_0})}{T_1}\right\rfloor\!-\!\left\lfloor\frac{T_2-d_1(t_{w_0})}{T_1}\right\rfloor$ can only be $0$ or $1$. We will use this result to simply equation (\ref{difference_1}) as follows.\vspace{0.5em}\\
1. If $\left\lfloor\!\frac{T_2-d_1'(t_{w_0})}{T_1}\!\right\rfloor\!-\!\left\lfloor\!\frac{T_2-d_1(t_{w_0})}{T_1}\!\right\rfloor\!=\!0$, then 
$\Delta_1\!=\!\!\min\!\left(T_1\!\left\{ \frac{T_2-d_1'(t_{w_0})}{T_1}\right\},C_1\right)\!-\!\min\!\left(T_1\!\left\{ \frac{T_2-d_1(t_{w_0})}{T_1}\right\},C_1\right)$. Because of $\frac{T_2-d_1'(t_{w_0})}{T_1}\!-\!\frac{T_2-d_1(t_{w_0})}{T_1}\!>\!0$ and $\left\lfloor\!\frac{T_2-d_1'(t_{w_0})}{T_1}\!\right\rfloor\!-\!\left\lfloor\!\frac{T_2-d_1(t_{w_0})}{T_1}\!\right\rfloor\!=\!0$, we can derive $\left\{\!\frac{T_2-d_1'(t_{w_0})}{T_1}\!\right\}-\left\{\!\frac{T_2-d_1(t_{w_0})}{T_1}\!\right\}\!=\!\frac{T_2-d_1'(t_{w_0})}{T_1}-\left\lfloor\!\frac{T_2-d_1'(t_{w_0})}{T_1}\!\right\rfloor-\left(\frac{T_2-d_1(t_{w_0})}{T_1}-\left\lfloor\!\frac{T_2-d_1(t_{w_0})}{T_1}\!\right\rfloor\right)\!=\!\frac{T_2-d_1'(t_{w_0})}{T_1}-\frac{T_2-d_1(t_{w_0})}{T_1}>0$. Therefore, we have the inequality $T_1\left\{\!\frac{T_2-d_1'(t_{w_0})}{T_1}\!\right\}\!>\!T_1\left\{\!\frac{T_2-d_1(t_{w_0})}{T_1}\!\right\}$. Because of the minimal operator, we discuss the following three cases:\\
1.a. If $T_1\left\{\!\frac{T_2-d_1'(t_{w_0})}{T_1}\!\right\}\!>\!T_1\left\{\!\frac{T_2-d_1(t_{w_0})}{T_1}\!\right\}\ge C_1$, then $\min\!\left(T_1\!\left\{\! \frac{T_2-d_1'(t_{w_0})}{T_1}\!\right\},C_1\!\right)\!=\!\min\!\left(\!T_1\left\{\! \frac{T_2-d_1(t_{w_0})}{T_1}\!\right\},C_1\!\right)\!=\!C_1$. Therefore, $\Delta_1\!=\!C_1\!-\!C_1\!=\!0$.\\
1.b. If $T_1\left\{\!\frac{T_2-d_1'(t_{w_0})}{T_1}\!\right\}\!>C_1>\!T_1\left\{\!\frac{T_2-d_1(t_{w_0})}{T_1}\!\right\}$, then $\min\!\left(T_1\!\left\{\! \frac{T_2-d_1'(t_{w_0})}{T_1}\!\right\},C_1\!\right)\!=\!C_1$ and $\min\!\left(\!T_1\left\{\! \frac{T_2-d_1(t_{w_0})}{T_1}\!\right\},C_1\!\right)\!=\!T_1\left\{\! \frac{T_2-d_1(t_{w_0})}{T_1}\right\}$. Therefore, we have $\Delta_1\!=\!C_1-T_1\left\{\! \frac{T_2-d_1(t_{w_0})}{T_1}\right\}\!>\!0$.\\
1.c. If $C_1\!\ge\!T_1\left\{\!\frac{T_2-d_1'(t_{w_0})}{T_1}\!\right\}\!>\!T_1\left\{\!\frac{T_2-d_1(t_{w_0})}{T_1}\!\right\}$, then $\min\!\left(T_1\!\left\{\! \frac{T_2-d_1'(t_{w_0})}{T_1}\!\right\},C_1\!\right)\!=\!T_1\!\left\{\! \frac{T_2-d_1'(t_{w_0})}{T_1}\!\right\}$ and $\min\!\left(\!T_1\left\{\! \frac{T_2-d_1(t_{w_0})}{T_1}\!\right\},C_1\!\right)\!=\!T_1\left\{\! \frac{T_2-d_1(t_{w_0})}{T_1}\right\}$. Therefore, we have $\Delta_1\!=\!T_1\!\left\{\! \frac{T_2-d_1'(t_{w_0})}{T_1}\!\right\}-T_1\left\{\! \frac{T_2-d_1(t_{w_0})}{T_1}\right\}\!>\!0$.\\
Summarizing the above three cases, we have $\Delta_1\!\ge\! 0$.\vspace{0.5em}\\
2. If $\left\lfloor\!\frac{T_2-d_1'(t_{w_0})}{T_1}\!\right\rfloor\!-\!\left\lfloor\!\frac{T_2-d_1(t_{w_0})}{T_1}\!\right\rfloor\!=\!1$, then 
$\Delta_1\!=\!C_1\!+\!\min\!\left(T_1\!\left\{ \frac{T_2-d_1'(t_{w_0})}{T_1}\right\},C_1\right)\!-\!\min\!\left(T_1\!\left\{ \frac{T_2-d_1(t_{w_0})}{T_1}\right\},C_1\right)$.
Because of the inequalities $\min\!\left(T_1\!\left\{ \frac{T_2-d_1'(t_{w_0})}{T_1}\right\},C_1\right)\!\ge\! 0$ and $\min\!\left(T_1\left\{ \frac{T_2-d_1(t_{w_0})}{T_1}\right\},C_1\right)\!\le\! C_1$, we have $\min\!\left(T_1\left\{ \frac{T_2-d_1'(t_{w_0})}{T_1}\right\},C_1\right)\!-\!\min\!\left(T_1\left\{ \frac{T_2-d_1(t_{w_0})}{T_1}\right\},C_1\right)\! \ge\! -C_1$ which leads to $\Delta_1\!\ge\! C_1\!-\!C_1\!=\!0$. \\
Summarizing {\bf Case $\bf 1$}, we can conclude that if $d_1(t_{w_0})$ is decreased, then ${\rm OP}_1(t_{w_0}, t_{w_0}\!+\!T_2, d_1(t_{w_0}))$ will increase or not change. Since $d_1(t_{w_0})\!>\!0$, we have ${\rm OP}_1(t_{w_0}, t_{w_0}\!+\!T_2, d_1(t_{w_0}))\!<\!{\rm OP}_1(t_{w_0}, t_{w_0}\!+\!T_2, 0)$, where ${\rm OP}_1(t_{w_0}, t_{w_0}\!+\!T_2, 0)\!=\!C_1\!\left\lfloor T_2/T_1\right\rfloor\!+\!\min\!\left(T_1\left\{ T_2/T_1\right\},C_1\right)$.\vspace{1em}\\
{\bf Case $\bf 2$} is $T_1\!-\!C_1\!<\! d_1(t_{w_0})\!\le\! T_1$. We will show that ${\rm OP}_1(t_{w_0}, t_{w_0}\!+\!T_2, d_1(t_{w_0}))$ is a non-decreasing function of $d_1(t_{w_0})$ in the range of $(T_1-C_1,T_1]$. Therefore, the maximum of ${\rm OP}_1(t_{w_0}, t_{w_0}\!+\!T_2, d_1(t_{w_0}))$ is obtained when $d_1(t_{w_0})=T_1$. Define a variable $d_1''(t_{w_0})$ which satisfies $T_1\!-\!C_1\!<\! d_1(t_{w_0})\!<\!d_1''(t_{w_0})\!\le\! T_1$ and its corresponding time-occupancy is denoted as ${\rm OP}_1(t_{w_0}, t_{w_0}\!+\!T_2, d_1''(t_{w_0}))$.  By taking the difference $\Delta_2\!=\!{\rm OP}_1(t_{w_0}, t_{w_0}\!+\!T_2, d_1''(t_{w_0}))\!-\!{\rm OP}_1(t_{w_0}, t_{w_0}\!+\!T_2, d_1(t_{w_0}))$, we will prove that $\Delta_2\ge 0$ is always satisfied for $T_1\!-\!C_1\!<\! d_1(t_{w_0})\!\le\! T_1$. 
Plugging in the second formula in (\ref{OP_1}), we have
\begin{align}\label{Difference_OP1_Case2}
\Delta_2&\!=\!d_1''(t_{w_0})\!-\!d_1(t_{w_0})\!+\!C_1\!\left(\!\left\lfloor\!\frac{T_2-d_1''(t_{w_0})}{T_1}\!\right\rfloor\!-\!\left\lfloor\!\frac{T_2-d_1(t_{w_0})}{T_1}\!\right\rfloor\!\right)\\
&\!+\!\min\!\left(T_1\!\left\{ \frac{T_2-d_1''(t_{w_0})}{T_1}\right\},C_1\right)\!-\!\min\!\left(T_1\!\left\{ \frac{T_2-d_1(t_{w_0})}{T_1}\right\},C_1\right).\nonumber
\end{align}
 The term $\left\lfloor\!\frac{T_2-d_1''1(t_{w_0})}{T_1}\!\right\rfloor\!-\!\left\lfloor\!\frac{T_2-d_1(t_{w_0})}{T_1}\!\right\rfloor$ can only be either $0$ or $-1$. The reasons are as follows. First, we analyze the values of $\frac{T_2-d_1''(t_{w_0})}{T_1}$ and $\frac{T_2-d_1(t_{w_0})}{T_1}$ by taking the difference $\frac{T_2-d_1''(t_{w_0})}{T_1}-\frac{T_2-d_1(t_{w_0})}{T_1}=\frac{d_1(t_{w_0})-d_1''(t_{w_0})}{T_1}<0$. Based on $T_1\!-\!C_1\!<\! d_1(t_{w_0})\!<\!d_1''(t_{w_0})\!\le\! T_1$, we have $-C_1<d_1(t_{w_0})\!-\!d_1''(t_{w_0})\!<\!0$, which leads to $-\frac{C_1}{T_1}\!<\!\frac{d_1(t_{w_0})-d_1''(t_{w_0})}{T_1}\!<\!0$. Since $C_1<T_1$, we can conclude that $-1\!<\!-\frac{C_1}{T_1}\!<\!\frac{T_2-d_1''(t_{w_0})}{T_1}\!-\!\frac{T_2-d_1(t_{w_0})}{T_1}\!<\!0$. The integer parts must satisfy $-1\le \left\lfloor\frac{T_2-d_1''(t_{w_0})}{T_1}\right\rfloor\!-\!\left\lfloor\frac{T_2-d_1(t_{w_0})}{T_1}\right\rfloor\le 0$, which only takes $0$ or $-1$. Similarly as the discussion of {\bf Case $\bf 1$}, we will use this result to simply equation (\ref{Difference_OP1_Case2}) as follows.\vspace{0.5em}\\
1. If $\left\lfloor\!\frac{T_2-d_1''1(t_{w_0})}{T_1}\!\right\rfloor\!-\!\left\lfloor\!\frac{T_2-d_1(t_{w_0})}{T_1}\!\right\rfloor\!=\!0$, then we have $\Delta_2\!=\!d_1''(t_{w_0})-d_1(t_{w_0})+\min\!\left(T_1\!\left\{ \frac{T_2-d_1''(t_{w_0})}{T_1}\right\},C_1\right)\!-\!\min\!\left(T_1\!\left\{ \frac{T_2-d_1(t_{w_0})}{T_1}\right\},C_1\right)$. We can also derive that $\left\{\! \frac{T_2-d_1''(t_{w_0})}{T_1}\!\right\}-\left\{\! \frac{T_2-d_1(t_{w_0})}{T_1}\!\right\}\!=\!\frac{T_2\!-\!d_1''(t_{w_0})}{T_1}-\frac{T_2\!-\!d_1(t_{w_0})}{T_1}\!=\!\frac{d_1(t_{w_0})\!-\!d_1''(t_{w_0})}{T_1}\!<\!0$, which means that $T_1\left\{\! \frac{T_2-d_1(t_{w_0})}{T_1}\!\right\}\!>\!T_1\left\{\! \frac{T_2-d_1''(t_{w_0})}{T_1}\!\right\}$.
Because of the minimal operator, we discuss the following three cases:\\
1.a. If $T_1\!\left\{\! \frac{T_2-d_1(t_{w_0})}{T_1}\!\right\}\!>\!T_1\!\left\{\! \frac{T_2-d_1''(t_{w_0})}{T_1}\!\right\}\!\ge\! C_1$, then $\min\!\left(T_1\!\left\{\! \frac{T_2-d_1''(t_{w_0})}{T_1}\!\right\},C_1\!\right)\!=\!\min\!\left(\!T_1\left\{\! \frac{T_2-d_1(t_{w_0})}{T_1}\!\right\},C_1\!\right)\!=\!C_1.$ Therefore, $\Delta_2\!=\!d_1''(t_{w_0})\!-\!d_1(t_{w_0})\!>\!0$.\\
1.b. If $T_1\!\left\{\! \frac{T_2-d_1(t_{w_0})}{T_1}\!\right\}\!>\!C_1\!>\!T_1\!\left\{\! \frac{T_2-d_1''(t_{w_0})}{T_1}\!\right\}$, then $\min\!\left(\!T_1\!\left\{\! \frac{T_2\!-\!d_1(t_{w_0})}{T_1}\!\right\}\!,C_1\!\right)\!=\!C_1$ and $\min\!\left(\!T_1\!\left\{\! \frac{T_2\!-\!d_1''(t_{w_0})}{T_1}\!\right\}\!,C_1\!\right)\!=\!T_1\!\left\{\! \frac{T_2\!-\!d_1''(t_{w_0})}{T_1}\!\right\}$. Therefore, $\Delta_2\!=\!d_1''(t_{w_0})\!-\!d_1(t_{w_0})\!+\!T_1\!\left\{\! \frac{T_2-d_1''(t_{w_0})}{T_1}\right\}\!-\!C_1\!\ge\! d_1''(t_{w_0})\!-\!d_1(t_{w_0})\!+\!T_1\!\left\{\! \frac{T_2-d_1''(t_{w_0})}{T_1}\!\right\}\!-\!T_1\!\left\{\! \frac{T_2-d_1(t_{w_0})}{T_1}\!\right\}$. Since $\left\{\! \frac{T_2-d_1''(t_{w_0})}{T_1}\!\right\}-\left\{\! \frac{T_2-d_1(t_{w_0})}{T_1}\!\right\}\!=\!\frac{d_1(t_{w_0})\!-\!d_1''(t_{w_0})}{T_1}$, we have $T_1\left\{\! \frac{T_2-d_1''(t_{w_0})}{T_1}\!\right\}-T_1\left\{\! \frac{T_2-d_1(t_{w_0})}{T_1}\!\right\}\!=\!d_1(t_{w_0})\!-\!d_1''(t_{w_0})$. Therefore, $\Delta_2\!\ge\!d_1''(t_{w_0})\!-\!d_1(t_{w_0})\!+\!d_1(t_{w_0})\!-\!d_1''(t_{w_0})=\!0$.\\
1.c. If $C_1\!\ge\!T_1\!\left\{\! \frac{T_2-d_1(t_{w_0})}{T_1}\!\right\}\!>\!T_1\!\left\{\! \frac{T_2-d_1''(t_{w_0})}{T_1}\!\right\}$, then $\min\!\left(\!T_1\!\left\{ \!\frac{T_2-d_1(t_{w_0})}{T_1}\!\right\},C_1\!\right)\!=\!T_1\!\left\{\! \frac{T_2-d_1(t_{w_0})}{T_1}\!\right\}$ and $\min\!\left(\!T_1\!\left\{\!\frac{T_2-d_1''(t_{w_0})}{T_1}\!\right\},C_1\!\right)\!=\!T_1\!\left\{\! \frac{T_2-d_1''(t_{w_0})}{T_1}\!\right\}$. Therefore, $\Delta_2\!=\!d_1''(t_{w_0})-d_1(t_{w_0})+T_1\!\left\{\! \frac{T_2-d_1''(t_{w_0})}{T_1}\!\right\}-T_1\!\left\{\! \frac{T_2-d_1(t_{w_0})}{T_1}\!\right\}\!=\!0$.\\
Summarizing the above three cases, we have $\Delta_2\!\ge\! 0$.\vspace{0.5em}\\
2. If $\left\lfloor\!\frac{T_2-d_1''1(t_{w_0})}{T_1}\!\right\rfloor\!-\!\left\lfloor\!\frac{T_2-d_1(t_{w_0})}{T_1}\!\right\rfloor\!=\!-1$, then we have  $\Delta_2\!=\!d_1''(t_{w_0})\!-\!d_1(t_{w_0})\!-\!C_1\!+\!\min\left(T_1\left\{ \frac{T_2-d_1''(t_{w_0})}{T_1}\right\},C_1\right)\!-\!\min\left(T_1\left\{ \frac{T_2-d_1(t_{w_0})}{T_1}\right\},C_1\right)$. In addition, the difference between the two fractional parts is $\left\{ \frac{T_2-d_1''(t_{w_0})}{T_1}\right\}-\left\{ \frac{T_2-d_1(t_{w_0})}{T_1}\right\}\!=\!\frac{T_2\!-\!d_1''(t_{w_0})}{T_1}-\frac{T_2\!-\!d_1(t_{w_0})}{T_1}-\left(\left\lfloor\frac{T_2-d_1''1(t_{w_0})}{T_1}\right\rfloor\!-\!\left\lfloor\frac{T_2-d_1(t_{w_0})}{T_1}\right\rfloor\right)\!=\!\frac{d_1(t_{w_0})-d_1''(t_{w_0})}{T_1}\!+\!1\!=\!\frac{d_1(t_{w_0})-d_1''(t_{w_0})+T_1}{T_1}$. Because $T_1\!-\!C_1\!<\! d_1(t_{w_0})\!<\!d_1''(t_{w_0})< T_1$, we can derive  $d_1(t_{w_0})\!-\!d_1''(t_{w_0})\!>\!-C_1$. Therefore, $\left\{\! \frac{T_2-d_1''(t_{w_0})}{T_1}\!\right\}\!-\!\left\{\! \frac{T_2-d_1(t_{w_0})}{T_1}\!\right\}\!>\!\frac{-C_1+T_1}{T_1}\!>\!0$, i.e., $\left\{\! \frac{T_2-d_1''(t_{w_0})}{T_1}\!\right\}\!>\!\left\{\! \frac{T_2-d_1(t_{w_0})}{T_1}\!\right\}$. Again, because of the minimal operator, we discuss the following three cases:\\
2.a. If we assume $T_1\!\left\{\! \frac{T_2-d_1''(t_{w_0})}{T_1}\!\right\}\!>\!T_1\!\left\{\! \frac{T_2-d_1(t_{w_0})}{T_1}\!\right\}\!\ge\! C_1$, i.e., $\left\{\! \frac{T_2-d_1(t_{w_0})}{T_1}\!\right\}\!\ge\! \frac{C_1}{T_1}$, then we have $\left\{\! \frac{T_2-d_1''(t_{w_0})}{T_1}\!\right\}\!>\!\frac{C_1}{T_1}\!+\!\frac{T_1-C_1}{T_1}\!=\!1$, which is contradict to the fact that $\left\{\! \frac{T_2-d_1''(t_{w_0})}{T_1}\!\right\}$ is a fractional part. Therefore, $T_1\!\left\{\! \frac{T_2-d_1''(t_{w_0})}{T_1}\!\right\}\!>\!T_1\!\left\{\! \frac{T_2-d_1(t_{w_0})}{T_1}\!\right\}\!\ge\! C_1$ will never occur.\vspace{1em}\\
2.b.  If $T_1\!\left\{\! \frac{T_2-d_1''(t_{w_0})}{T_1}\!\right\}\!>\!C_1\!>\! T_1\!\left\{\! \frac{T_2-d_1(t_{w_0})}{T_1}\!\right\}$, then $\min\!\left(\!T_1\!\left\{\! \frac{T_2-d_1''(t_{w_0})}{T_1}\!\right\},C_1\!\right)\!=\!C_1$ and $\min\!\left(\!T_1\!\left\{\! \frac{T_2-d_1(t_{w_0})}{T_1}\!\right\}\!,C_1\!\right)\!=\!T_1\!\left\{\! \frac{T_2-d_1(t_{w_0})}{T_1}\!\right\}$. Therefore, we have $\Delta_2\!=\!d_1''(t_{w_0})\!-\!d_1(t_{w_0})\!-\!C_1\!+\!C_1\!-\!T_1\!\left\{\! \frac{T_2-d_1(t_{w_0})}{T_1}\!\right\}\!=\! d_1''(t_{w_0})\!-\!d_1(t_{w_0})\!-\!T_1\!\left\{\! \frac{T_2-d_1(t_{w_0})}{T_1}\!\right\}$. We can rewrite $d_1''(t_{w_0})\!-\!d_1(t_{w_0})$ as $T_1\frac{T_2-d_1(t_{w_0})}{T_1}-T_1\frac{T_2-d_1''(t_{w_0})}{T_1}$, which leads to  $\Delta_2\!=\!T_1\frac{T_2-d_1(t_{w_0})}{T_1}-T_1\frac{T_2-d_1''(t_{w_0})}{T_1}-T_1\!\left\{\frac{T_2-d_1(t_{w_0})}{T_1}\right\}$. If we reorder the second and third term, we have $\Delta_2\!=\!\left(T_1\frac{T_2-d_1(t_{w_0})}{T_1}-T_1\!\left\{\frac{T_2-d_1(t_{w_0})}{T_1}\right\}\right)-T_1\frac{T_2-d_1''(t_{w_0})}{T_1}$, which can be rewritten as $T_1\!\left\lfloor\frac{T_2-d_1(t_{w_0})}{T_1}\right\rfloor\!-\!T_1\frac{T_2-d_1''(t_{w_0})}{T_1}$. We can also rewrite $T_1\frac{T_2-d_1''(t_{w_0})}{T_1}$ as $\left(T_1\!\left\lfloor\frac{T_2-d_1''(t_{w_0})}{T_1}\right\rfloor\!+\!T_1\!\left\{\frac{T_2-d_1''(t_{w_0})}{T_1}\right\}\right)$. Therefore, the difference  $\Delta_2\!=\!T_1\!\left\lfloor\frac{T_2-d_1(t_{w_0})}{T_1}\right\rfloor-T_1\!\left\lfloor\frac{T_2-d_1''(t_{w_0})}{T_1}\right\rfloor\!-\!T_1\!\left\{\frac{T_2-d_1''(t_{w_0})}{T_1}\right\}$. Since $\left\lfloor\!\frac{T_2-d_1''1(t_{w_0})}{T_1}\!\right\rfloor\!-\!\left\lfloor\!\frac{T_2-d_1(t_{w_0})}{T_1}\!\right\rfloor\!=\!-1$, we have $T_1\left\lfloor\!\frac{T_2-d_1(t_{w_0})}{T_1}\!\right\rfloor\!-\!T_1\left\lfloor\!\frac{T_2-d_1''(t_{w_0})}{T_1}\!\right\rfloor\!=\!T_1$ and $\Delta_2\!=\!T_1\!-\!T_1\!\left\{\frac{T_2-d_1''(t_{w_0})}{T_1}\right\}\!=\!T_1\left(1-\left\{\frac{T_2-d_1''(t_{w_0})}{T_1}\right\}\right)$. And because $\left\{\frac{T_2-d_1''(t_{w_0})}{T_1}\right\}$ is the fractional part, which must be within the range $[0,1)$, we have the result $\Delta_2\!=\!T_1\left(1-\left\{\frac{T_2-d_1''(t_{w_0})}{T_1}\right\}\right)\!>\!0$.\\
2.c. If $C_1\!\ge\!T_1\!\left\{\! \frac{T_2-d_1''(t_{w_0})}{T_1}\!\right\}\!>\!T_1\!\left\{\! \frac{T_2-d_1(t_{w_0})}{T_1}\!\right\}$, then $\min\!\left(\!T_1\!\left\{\! \frac{T_2-d_1''(t_{w_0})}{T_1}\!\right\},C_1\!\right)\!=\!T_1\!\left\{ \!\frac{T_2-d_1''(t_{w_0})}{T_1}\!\right\}$ and 
$\min\!\left(T_1\!\left\{\! \frac{T_2-d_1(t_{w_0})}{T_1}\!\right\},C_1\right)\!=\!T_1\!\left\{\! \frac{T_2-d_1(t_{w_0})}{T_1}\!\right\}$. Therefore, we have $\Delta_2\!=\!d_1''(t_{w_0})\!-\!d_1(t_{w_0})\!-\!C_1\!+\!T_1\!\left\{\! \frac{T_2-d_1''(t_{w_0})}{T_1}\!\right\}\!-\!T_1\!\left\{\! \frac{T_2-d_1(t_{w_0})}{T_1}\!\right\}$. Because $\left\{ \frac{T_2-d_1''(t_{w_0})}{T_1}\right\}\!-\!\left\{ \frac{T_2-d_1(t_{w_0})}{T_1}\right\}\!=\!\frac{d_1(t_{w_0})-d_1''(t_{w_0})+T_1}{T_1}$, we have $T_1\left\{ \frac{T_2-d_1''(t_{w_0})}{T_1}\right\}\!-\!T_1\left\{ \frac{T_2-d_1(t_{w_0})}{T_1}\right\}\!=\!d_1(t_{w_0})-d_1''(t_{w_0})+T_1$. Therefore,  $\Delta_2\!=\!d_1''(t_{w_0})\!-\!d_1(t_{w_0})\!-\!C_1\!+\!d_1(t_{w_0})-d_1''(t_{w_0})+T_1\!=\!-C_1\!+\!T_1\!>\!0$.\\
Summarizing the above three cases, we have $\Delta_2\!\ge\! 0$.
Summarizing {\bf Case $\bf 2$}, we can conclude that if $d_1(t_{w_0})$ is increased, then ${\rm OP}_1(t_{w_0}, t_{w_0}\!+\!T_2, d_1(t_{w_0}))$ will increase or not change. The maximal value is obtained when $d_1^*(t_{w_0})\!=\!T_1$ and ${\rm OP}_{\rm max}\!=\!{\rm OP}_1(t_{w_0}, t_{w_0}\!+\!T_2, T_1)\!=\!C_1\left\lfloor T_2/T_1\right\rfloor\!+\!\min\left(T_1\left\{ T_2/T_1\right\},C_1\right)$.

Hence, combining {\bf Case $\bf 1$} and {\bf Case $\bf 2$}, the maximal value of ${\rm OP}_1(t_{w_0}, t_{w_0}\!+\!T_2, d_1(t_{w_0}))$ is obtained when $d_1^*(t_{w_0})\!=\!T_1$. 

\subsection{Proof of Corollary \ref{Coro_worst_case_moments}}\label{Proof_corollary_worst_moment}
In the proof of \textbf{Theorem \ref{lemma2}}, under the discussion $1$.c of {\bf Case $\bf 2$}, we have $\Delta_2\!=\!0$. It means that if there exists a deadline variable $d_1(t_{w_0})$ satisfying three conditions in $1$.c of {\bf Case $\bf 2$}, i.e., $T_1\!-\!C_1\!<\! d_1(t_{w_0})\!<\! T_1$, $\left\lfloor\!\frac{T_2-d_1(t_{w_0})}{T_1}\!\right\rfloor\!=\!\left\lfloor\!\frac{T_2-T_1}{T_1}\!\right\rfloor$ and $C_1\!\ge\!T_1\!\left\{\! \frac{T_2-d_1(t_{w_0})}{T_1}\!\right\}\!\ge\!T_1\!\left\{\! \frac{T_2-T_1}{T_1}\!\right\}$, then ${\rm OP}_1(t_{w_0}, t_{w_0}\!+\!T_2, d_1(t_{w_0}))\!=\!{\rm OP}_1(t_{w_0}, t_{w_0}\!+\!T_2, T_1)\!=\!{\rm OP}_{\rm max}$ with $d_1(t_{w_0})\!\neq\! T_1$. In other words, all $d_1(t_{w_0})$ which satisfy the three conditions under $1$.c of {\bf Case $\bf 2$} are the worst-case deadlines. 

To satisfy these three conditions, the timing characteristics $T_1$, $T_2$ and $C_1$ first need to satisfy $C_1\!\ge\!T_1\!\left\{\! \frac{T_2-T_1}{T_1}\!\right\}$, which can be represented as a requirement for $T_2$ such that $MT_1\!\le\!T_2\!\le\!MT_1\!+\!C_1$ for some integer $M\!\ge\!1$. With this new representation, we have $\left\lfloor\!\frac{T_2-T_1}{T_1}\!\right\rfloor\!=\!M\!-\!1$ and $\left\lfloor\!\frac{T_2-d_1(t_{w_0})}{T_1}\!\right\rfloor\!=\!\left\lfloor\!M\!+\!\left\{\frac{T_2}{T_1}\right\}-\frac{d_1(t_{w_0})}{T_1}\!\right\rfloor\!=\!M\!-\!1$. Hence, we have $-1\!\le\!\left\{\frac{T_2}{T_1}\right\}-\frac{d_1(t_{w_0})}{T_1}\!<\!0$, which leads to $T_1\!\left\{\frac{T_2}{T_1}\right\}\!<\!d_1(t_{w_0})\!\le\!T_1\!\left\{\frac{T_2}{T_1}\right\}\!+\!T_1$.
As for the last condition $C_1\!\ge\!T_1\!\left\{\! \frac{T_2-d_1(t_{w_0})}{T_1}\!\right\}\!\ge\!T_1\!\left\{\! \frac{T_2-T_1}{T_1}\!\right\}$, leveraging the facts that $\left\{\! \frac{T_2-d_1(t_{w_0})}{T_1}\!\right\}\!=\! \frac{T_2-d_1(t_{w_0})}{T_1}\!-\!\left\lfloor\! \frac{T_2-d_1(t_{w_0})}{T_1}\!\right\rfloor$ and $\left\lfloor\!\frac{T_2-d_1(t_{w_0})}{T_1}\!\right\rfloor\!=\!M\!-\!1$, we have  $T_1\!\left\{\! \frac{T_2-d_1(t_{w_0})}{T_1}\!\right\}\!=\!T_2\!-\!d_1(t_{w_0})\!-\!(M\!-\!1)T_1\!=\!T_1\!+\!T_1\!\left\{\! \frac{T_2}{T_1}\!\right\}\!-\!d_1(t_{w_0})\!\le\!C_1$, which leads to $T_1\!-\!C_1\!+\!T_1\!\left\{\! \frac{T_2}{T_1}\!\right\}\!\le\!d_1(t_{w_0})$. Combining the range requirements of $d_1(t_{w_0})$, we have $T_1\!-\!C_1\!+\!T_1\!\left\{ \frac{T_2}{T_1}\right\}\!\le\! d_1(t_{w_0})\!\le\!T_1$ with $MT_1\!\le\!T_2\!\le\!MT_1\!+\!C_1$ for some integer $M\!\ge\!1$. Specially, when $T_2\!=\!MT_1\!+\!C_1$, we have $T_1\!-\!C_1\!+\!T_1\!\left\{ \frac{T_2}{T_1}\right\}\!=\!T_1$ and $T_1\!\le\! d_1(t_{w_0})\!\le\!T_1$, which makes $d_1(t_{w_0})\!=\!T_1$ the only worst-case deadline.

Similarly, in the discussion under $1$.a of {\bf Case $\bf 1$}, if there exists a deadline variable $d_1(t_{w_0})$ satisfying $0\!<\! d_1(t_{w_0})\!\le\!T_1\!-\!C_1$, $\left\lfloor\!\frac{T_2-d_1(t_{w_0})}{T_1}\!\right\rfloor\!=\!\left\lfloor\!\frac{T_2}{T_1}\!\right\rfloor$ and $\!T_1\!\left\{\! \frac{T_2-d_1(t_{w_0})}{T_1}\!\right\}\!\ge\!T_1\!\left\{\! \frac{T_2}{T_1}\!\right\}\!\ge\! C_1$,  we have $\Delta_1\!=\!0$, which means ${\rm OP}_1(t_{w_0}, t_{w_0}\!+\!T_2, d_1(t_{w_0}))\!=\!{\rm OP}_1(t_{w_0}, t_{w_0}\!+\!T_2, 0)\!=\!{\rm OP}_{\rm max}$. In this case, the timing characteristics need to satisfy $T_1\!\left\{\! \frac{T_2}{T_1}\!\right\}\!\ge\! C_1$, which can be represented as a requirement for $T_2$ such that $MT_1\!+\!C_1\!\le\!T_2\!<\!(M\!+\!1)T_1$ for some integer $M\!\ge\!1$. And from $\left\lfloor\!\frac{T_2-d_1(t_{w_0})}{T_1}\!\right\rfloor\!=\!\left\lfloor\!\frac{T_2}{T_1}\!\right\rfloor\!=\!M$, we have $\left\lfloor\!\frac{T_2-d_1(t_{w_0})}{T_1}\!\right\rfloor\!=\!\left\lfloor\!M\!+\!\left\{\frac{T_2}{T_1}\right\}\!-\!\frac{d_!(t_{w_0})}{T_1}\!\right\rfloor\!=\!M$, which leads to $0\!\le\!\left\{\frac{T_2}{T_1}\right\}-\frac{d_1(t_{w_0})}{T_1}\!<\!1$, i.e., $T_1\!\left\{\frac{T_2}{T_1}\right\}\!-\!T_1\!<\!d_1(t_{w_0})\!\le\! T_1\!\left\{\frac{T_2}{T_1}\right\}$. As for the last condition $\!T_1\!\left\{\! \frac{T_2-d_1(t_{w_0})}{T_1}\!\right\}\!\ge\! C_1$, we have $T_1\!\left\{\! \frac{T_2-d_1(t_{w_0})}{T_1}\!\right\}\!=\!T_2\!-\!d_1(t_{w_0})\!-\!MT_1\!=\!T_1\!\left\{\! \frac{T_2}{T_1}\!\right\}\!-\!d_1(t_{w_0})\!\ge\!C_1$, which leads to $d_1(t_{w_0})\!\le\!T_1\left\{\! \frac{T_2}{T_1}\!\right\}\!-\!C_1$. Combining the range requirements of $d_1(t_{w_0})$ in this case, we have $0\!<\! d_1(t_{w_0})\!\le\!T_1\!\left\{ \frac{T_2}{T_1}\right\}\!-\!C_1$ with $MT_1\!+\!C_1\!\le\!T_2\!<\!(M\!+\!1)T_1$ for some integer $M\!\ge\!1$. Also, when $T_2\!=\!MT_1\!+\!C_1$, we have $T_1\!\left\{ \frac{T_2}{T_1}\right\}\!-\!C_1\!=\!0$ and $0\!<\! d_1(t_{w_0})\!\le\!0$, where $d_1(t_{w_0})$ does not exist.

Specially, when $T_2\!=\!MT_1$, the time-occupancy ${\rm OP}_1(t_{w_0}, t_{w_0}\!+\!T_2, d_1(t_{w_0}))$ remain the same for all $0<d_1(t_{w_0})\!\le\!T_1$.
\subsection{Proof of Theorem \ref{lemma_max}}\label{Proof_theorem_2}
For Case $1$ in (\ref{OP_2}), if $T_1\!\le\!C_2$, then $\min\!\left(T_1\!-\!d_2(t_{w_0}'), C_2\right)\!=\!T_1\!-\!d_2(t_{w_0}')$. And since $0\!<\! d_2(t_{w_0}')\!\le\! T_2\!-\!C_2$, then the supreme of $T_1\!-\!d_2(t_{w_0}')$ is $T_1$ which is less than $C_2$. If $T_1\!>\! C_2$, then the term $T_1\!-\!d_2(t_{w_0}')$ can be larger than $C_2$ when $d_2(t_{w_0}')$ is very small. Therefore, ${\rm OP}_2(t_{w_0}', t_{w_0}'\!+\!T_1, d_2(t_{w_0}'))\!=\!\min\!\left(T_1\!-\!d_2(t_{w_0}'), C_2\right)\!=\!C_2$ when $0\!<\! d_2(t_{w_0}')\! \le \!T_1\!-\!C_2$.\\
For Case $2$, ${\rm OP}_2(t_{w_0}', t_{w_0}'\!+\!T_1, d_2(t_{w_0}'))$ is a constant $T_1\!-\!T_2\!+\!C_2$. Since $T_1 \!<\! T_2$, we have ${\rm OP}_2(t_{w_0}', t_{w_0}'\!+\!T_1, d_2(t_{w_0}'))\!=\!T_1\!-\!T_2\!+\!C_2\! <\! C_2$.\\
For Case $3$, $d_2(t_{w_0}')\!-\!(T_2\!-\!C_2)$ increases if $d_2(t_{w_0}')$ increases. Therefore, the maximal value of ${\rm OP}_2(t_{w_0}', t_{w_0}'\!+\!T_1, d_2(t_{w_0}'))$ equals $C_2$ when $d_2(t_{w_0}')\!=\!T_2$.
\subsection{Proof of Theorem \ref{theorem_N_system}}\label{Proof_theorem_N_system}

In this proof, we use the notation ${\rm OP}_q'$ to represent the time-occupancy corresponding to $d_q'(t_{w_0})\!=\!T_q$ and ${\rm OP}_q$ corresponding to $d_q(t_{w_0})\!\neq\!T_q$. We will prove this theorem by showing that, if there exists an arbitrary system $q\!<\!i$ with $d_q(t_{w_0})\!\neq\!T_q$, then we will compare the time-occupancy corresponding to $d_q(t_{w_0})$ with the time-occupancy corresponding to $d_q'(t_{w_0})\!=\!T_q$ and show that $\sum_{j=q}^{i-1}{\rm OP}_j'(t_{w_0}, t_{w_0}\!+\!T_i, d_j'(t_{w_0}), r_j'(t_{w_0}))\!\ge\! \sum_{j=q}^{i-1}{\rm OP}_j(t_{w_0}, t_{w_0}\!+\!T_i, d_j(t_{w_0}), r_j(t_{w_0}))$ from $q\!=\!i\!-\!1$ to $q\!=\!1$ by mathematical induction. 

First, we consider system $i\!-\!1$. If $d_{i-1}(t_{w_0})\!\neq\! T_{i-1}$, then we also consider the case where $d_{i-1}'(t_{w_0})=T_{i-1}$. For systems $1$ to $i\!-\!2$, since they have higher priorities than system $\!i\!-\!1$, their time-occupancies are identical under the two cases. Therefore, we only need to compare the values of ${\rm OP}_{i-1}(t_{w_0}, t_{w_0}\!+\!T_i, d_{i-1}(t_{w_0}), r_{i-1}(t_{w_0}))$ and ${\rm OP}_{i-1}'(t_{w_0}, t_{w_0}\!+\!T_i, d_{i-1}'(t_{w_0}), r_{i-1}'(t_{w_0}))$. Between systems $i\!-\!1$ and $i$, system $i\!-\!1$ has higher priority, which is the same as system $1$ in the two systems case we discussed in Section \ref{Section_two_systems_1}. Therefore, we can show that ${\rm OP}_{i-1}'(t_{w_0}, t_{w_0}\!+\!T_i, d_{i-1}'(t_{w_0}), r_{i-1}'(t_{w_0}))$ obtains its maximal value at the worst-case deadline using the same argument from proof in Appendix \ref{Proof_theorem_1} with $i\!-\!1$ replacing $1$. Hence, we have the inequality ${\rm OP}_{i-1}'(t_{w_0}, t_{w_0}\!+\!T_i, d_{i-1}'(t_{w_0}), r_{i-1}'(t_{w_0}))\!\ge\!{\rm OP}_{i-1}(t_{w_0}, t_{w_0}\!+\!T_i, d_{i-1}(t_{w_0}), r_{i-1}(t_{w_0}))$.

Then, assuming the deadline variable $d_{i-1}(t_{w_0})$ has been set to be $T_{i-1}$. Let us consider system $i\!-\!2$. If $d_{i-2}(t_{w_0})\!\neq\! T_{i-2}$, then we also consider the case where $d_{i-2}'(t_{w_0})\!=\!T_{i-2}$. Similarly as the case for system $i\!-\!1$, the time-occupancies of systems $1$ to $i\!-\!3$ are identical under these two cases because they have higher priority than system $i\!-\!2$. Among systems $i\!-\!2$, $i\!-\!1$ and $i$, system $i\!-\!2$ has highest priority. Therefore, we can easily show that ${\rm OP}_{i-2}'(t_{w_0}, t_{w_0}\!+\!T_i, d_{i-2}'(t_{w_0}), r_{i-2}'(t_{w_0}))$ is the maximal time-occupancy because $d_{i-2}'(t_{w_0})\!=\!T_{i-2}$ is the worst-case deadline. However, if the time-occupancy of system $i\!-\!2$ under $d_{i-2}'(t_{w_0})$ is larger, then the time-occupancy of system $i\!-\!1$ under $d_{i-2}'(t_{w_0})$ may be smaller than time-occupancy of system $i\!-\!1$ under $d_{i-2}(t_{w_0})$ because more time may be occupied by system $i\!-\!2$ under $d_{i-2}'(t_{w_0})$. Based on {\bf Lemma \ref{lemma_principle}}, the reduced time-occupancy from a lower prioritized system must all be occupied by the higher prioritized system. Therefore, the decreased time-occupancy of system $i\!-\!1$ should be less than or equal to the increased time-occupancy of system $i\!-\!2$. Hence, we have $\sum_{j=i-2}^{i-1}{\rm OP}_{j}'(t_{w_0}, t_{w_0}\!+\!T_i, T_j, 0)\!\ge\!\sum_{j=i-2}^{i-1}{\rm OP}_{j}(t_{w_0}, t_{w_0}\!+\!T_i, d_j(t_{w_0}), r_j(t_{w_0}))$.

For a system $q$ where $1\!\le\! q\!\le\!i\!-\!2$, we assume that the deadline variables of all lower prioritized systems $j\!=\!q\!+\!1$ to $i\!-\!1$ have been set to be $T_j$. The time-occupancies of systems $1$ to $q\!-\!1$ are identical under the two cases $d_{q}(t_{w_0})\!\neq\! T_{q}$ and $d_{q}'(t_{w_0})\!=\! T_{q}$. Therefore, system $q$ has highest priority among systems $j\!=\!q\!+\!1$ to $i\!-\!1$ and ${\rm OP}_{q}'(t_{w_0}, t_{w_0}\!+\!T_i, d_{q}'(t_{w_0}), r_{q}'(t_{w_0}))$ is the maximal time-occupancy. For any lower prioritized system $m\!=\!q\!+\!1,...,i\!-\!1$, if its time-occupancy is reduced under the case $d_q'(t_{w_0})$, the reduced amount of time must all be occupied by the higher prioritized systems. The decreased time-occupancy of system $m$ should be less than or equal to the increased time-occupancy of system $q$. Therefore, the summation of the time-occupancies will not decrease, which leads to the result $\sum_{j=q}^{i-1}{\rm OP}_{j}'(t_{w_0}, t_{w_0}\!+\!T_i, T_j, 0)\!\ge\!\sum_{j=q}^{i-1}{\rm OP}_{j}(t_{w_0}, t_{w_0}\!+\!T_i, d_j(t_{w_0}), r_j(t_{w_0}))$ for $q=1,...,i-2$.

Then repeating the process from $q\!=\!i\!-\!1$ to $1$, we can obtain the maximal value of $\sum_{j=1}^{i-1}{\rm OP}_j(t_{w_0}, t_{w_0}\!+\!T_i, d_j(t_{w_0}), r_j(t_{w_0}))$ when $d_j^*(t_{w_0})\!=\!T_j$ for all $j$. At the time $d_j^*(t_{w_0})\!=\!T_j$, the remaining time variable $r_j^*(t_{w_0})\!=\!C_j$.
\subsection{Proof of Theorem \ref{theorem_2_bound}}\label{Proof_theorem_bound}
Let systems $1$ and $2$ have periods $T_1$ and $T_2$, respectively. Assume that $T_1\!\le\! T_2$. According to RMS, system $1$ has higher priority. We will adjust $C_2$ to fully utilize the available shared resource time. Because of the minimal operator in (\ref{Theorem_1_OP1_formula}), we need to consider two cases:\\
Case $1$: if $C_1\!\le\! T_1\!\left\{T_2/T_1\right\}$, then $\min\! \left(T_1\left\{T_2/T_1\right\},C_1\right)\!=\!C_1$. Therefore, $C_2\!\le\! T_2\!-\!C_1\!\left\lfloor T_2/T_1\right\rfloor \!-\!C_1\!=\!T_2\!-\!C_1\!\left\lceil T_2/T_1\right\rceil$ where $\lceil\cdot \rceil$ is the rounding up operator. Then $U\!=\!\frac{C_1}{T_1}\!+\!\frac{C_2}{T_2}\!\le\! 1\!+\!C_1\!\left[(1/T_1)-(1/T_2)\lceil T_2/T_1\rceil\right]$ and $U$ decreases if $C_1$ increases.\\
Case $2$: if $C_1>T_1\left\{T_2/T_1\right\}$, then $\min \left(T_1\left\{T_2/T_1\right\},C_1\right)=T_1\left\{T_2/T_1\right\}$. Therefore, $C_2\le T_2-C_1\left\lfloor T_2/T_1\right\rfloor -T_1\left\{T_2/T_1\right\}=-C_1\left\lfloor T_2/T_1\right\rfloor+T_2\left\lfloor T_2/T_1\right\rfloor$. Then $U=\frac{C_1}{T_1}+\frac{C_2}{T_2}\le(T_2/T_1)\lfloor T_2/T_1\rfloor+C_1\left[(1/T_1)-(1/T_2)\lfloor T_2/T_1\rfloor\right]$ and $U$ increases if $C_1$ increases.\\
We then obtain the upper bound of the utilization factor as
\begin{align}
\overline{U}\!\left(\!C_1,\frac{T_2}{T_1}\!\right)\!=\!\!
\begin{cases} 1\!+\!C_1\!\left[(1/T_1)\!-\!(1/T_2)\lceil T_2/T_1\rceil\right], \text{if }0\! \le \! C_1 \! \le \! T_1\{T_2/T_1\}\\ 
(T_2/T_1)\!\lfloor T_2/T_1\rfloor\!+\!C_1\!\left[(1/T_1)\!-\!(1/T_2)\lfloor T_2/T_1\rfloor\right]
, \text{if }C_1 \!>\! T_1\{T_2/T_1\} \end{cases}\nonumber
\end{align}
When $C_1 \!=\!T_1\{T_2/T_1\}$, we have
\begin{align}\label{U_min}
\overline{U}\!\left(\!T_1\!\left\{\frac{T_2}{T_1}\right\},\frac{T_2}{T_1}\!\right)\!=\!1\!-\!(T_1/T_2)\left[\lceil T_2/T_1 \rceil\!-\!(T_2/T_1) \right]\left[(T_2/T_1)\!-\!\lfloor T_2/T_1 \rfloor \right]    
\end{align}
Let $I\!=\!\lfloor T_2/T_1 \rfloor$ and $f\!=\!\{T_2/T_1\}$, we can rewrite (\ref{U_min}) as 
$$\overline{U}(I,f)\!=\!1\!-\!f(1-f)/(I+f).$$
Since $\overline{U}(I,f)$ is monotonic increasing with $I$, the minimum of $\overline{U}(I,f)$ occurs at the smallest possible value of $I$, namely, $I\!=\!1$. Then when minimizing $U$ over $f$, we can take the derivative of $\overline{U}(1,f)$ with respect to $f$ and have
\begin{align}
\frac{\partial \overline{U}(I,f)}{\partial f}\!=\!\frac{f^2+2f-1}{(1+f)^2}.\nonumber
\end{align}
When $f\!=\!\sqrt{2}\!-\!1$, $\frac{\partial \overline{U}_{\rm min}(1,f)}{\partial f}\!=\!0$. And if $0 \!\le\! f\!<\!\sqrt{2}\!-\!1$, $\frac{\partial \overline{U}_{\rm min}(1,f)}{\partial f}\!<\!0$ and if $\sqrt{2}\!-\!1<f<1$,$\frac{\partial \overline{U}_{\rm min}(1,f)}{\partial f}\!>\!0$, $\overline{U}_{\rm}\!=\!\overline{U}(2,\sqrt{2}\!-\!1)=2(\sqrt{2}\!-\!1)$, which is the relation we want to prove. 

\bibliographystyle{spbasic}      
\bibliography{main}   

\end{document}